\documentclass{article}

\usepackage{arxiv}

\usepackage[utf8]{inputenc} 
\usepackage[T1]{fontenc}    
\usepackage{hyperref}       
\usepackage{url}            
\usepackage{booktabs}       
\usepackage{amsfonts}       
\usepackage{nicefrac}       
\usepackage{microtype}      
\usepackage{lipsum}		
\usepackage{graphicx}
\usepackage{doi}

\usepackage{amsmath}
\usepackage{amssymb}
\usepackage{multirow}
\usepackage{mathtools}
\usepackage[flushleft]{threeparttable}
\usepackage{booktabs,caption}
\usepackage{float}
\usepackage{graphicx}
\usepackage{lscape}

\makeatletter
\renewcommand\@biblabel[1]{}
\makeatother

\title{Variable selection in frailty mixture cure models via penalized likelihood estimation}

\author{ {Richard Tawiah}\thanks{Corresponding author.} \\
	School of Public Health and Preventive Medicine\\
	Monash University\\
	Melbourne, VIC 3004, Australia \\
	\texttt{richard.tawiah@monash.edu} \\
	\And
	{Shu Kay Ng} \\
	School of Medicine and Dentistry\\
	Griffith University\\
	Brisbane, QLD 4111, Australia \\
	\texttt{s.ng@griffith.edu.au} \\
    \And
{Geoffrey J.~McLachlan} \\
	School of Mathematics and Physics\\
	University of Queensland\\
	Brisbane, QLD 4072, Australia \\
	\texttt{g.mclachlan@uq.edu.au} \\
}

\renewcommand{\shorttitle}{Penalized REML estimation in frailty cure models}

\hypersetup{
pdftitle={A template for the arxiv style},
pdfsubject={q-bio.NC, q-bio.QM},
pdfauthor={Richard Tawiah, Shu Kay Ng, Geoffrey J.~McLachlan},
pdfkeywords={Adaptive lasso, EM algorithm, Generalized linear mixed model,  frailty mixture cure model, penalized variable selection, random effect, recurrent event data, regularization, REML, SCAD},
}

\begin{document}
\maketitle

\begin{abstract}
Variable selection naturally arises as a useful subject when faced with data with massive predictor space. In addition to the massive dimensionality, the data may be characterized by intra-subject correlation, and cure fraction, which are ubiquitous in longitudinal studies with recurrent events defining the endpoint of interest. However, variable selection methods simultaneously adjusting for intra-subject correlation, and cure fraction are rare. We propose a comprehensive variable selection method for frailty mixture cure models based on penalized least squares approximation via the generalized linear mixed model methodology. The method provides shrinkage estimation and selection of fixed effects in the incidence and the latency submodels, adjusting for intra-subject correlation using a random effect term. The random effect is shared between the incidence and the latency, incorporating a flexible choice of covariance structure, allowing intra-subject correlation to be modeled as either time-invariant or time-varying. Estimation is facilitated by a penalized semiparametric restricted maximum likelihood method using an expectation--maximization algorithm. Two penalty functions, namely the adaptive least absolute shrinkage and selection operator (adaptive lasso), and the smoothly clipped absolute deviation (SCAD) are studied in the proposed method.  Simulation studies are considered, benchmarking the method against an oracle procedure to access its finite sample performance. The practical utility of the method is illustrated using data on recurrent events from a breast cancer gene expression study. In the presence of a relatively large predictor space, results show that the method yields plausible interpretability in whole, as opposed to an unpenalized model. 
\end{abstract}


\section{Introduction}
Longitudinal studies often provide data on recurrent events such as sequential episodes of asthma attacks, multiple emergency department visits, and repeated occurrences of bladder cancer tumors. In biomedical research, these data provide important measurements for investigating the etiology of progressive diseases, and are usually considered as useful endpoints in clinical trials for assessing treatment benefits.  Naive statistical approaches to the analysis of recurrent events data focus on modeling the first events per subject, while neglecting subsequent ones. These approaches fall short because they do not utilize the complete information in the data. Multivariate methods provide effective techniques for modeling the data, taking advantage of the full information to increase statistical power, while adjusting for the inherent intra-subject correlation to allow a valid statistical inference. Intra-subject correlation arises because the occurrence of initial disease episodes may induce a biological phenomenon that modifies the risk for further occurrences. This may typically occur, for example, in heart disease studies where patients suffer one or more episodes of myocardial infarction, leading to progressive weakening of the heart muscle and hence increased risk for further recurrence. In addition to intra-subject correlation, a cure fraction may be present in the data, which is a characterized feature of individuals who get cured and remain recurrence-free, due to the effectiveness of treatment or other factors. For example, in clinical trials of cancer, certain novel therapeutic agents such as immunotherapy and targeted therapies have demonstrated the capacity in prolonging survival and providing protection against recurrence, compared to standard treatments such as chemotherapy (Felizzi et al. 2021). In practice, recurrence-free survival over a considerable period of time is often considered a possible cure, particularly for highly treatable diseases. Neglecting the existence of cure fraction in statistical analysis leads to overestimation of the survival function of uncured patients (Corbi\`ere et al. 2009). Frailty mixture cure models (Yu 2008; Rondeau et al. 2011; Tawiah et al. 2020a; Tawiah et al. 2020b) were developed specifically for recurrent events data, incorporating random effects (or frailty) and logistic regression model into standard survival models to concurrently adjust for cure fraction, and intra-subject correlation. These models afford several practical benefits. Nevertheless, associated methodological approaches for performing variable selection are rare, possibly due to the formidable estimation challenges posed by the intractable integration in the underlying marginal likelihood.  Moreover, these models are heavily parameterized than ordinary regression models, and so interpretation can be messy when faced with a massive predictor space. Therefore, variable selection methods are critical therein to improve model interpretability and prediction accuracy.

Variable selection by penalization methods is increasingly gaining popularity. Compared to traditional methods (e.g., subset, and stepwise selection), they afford relatively less computational cost and do not suffer instability issues. Furthermore, their sampling properties are data-dependent and easy to establish as opposed to the traditional methods (Fan and Li 2001). 
Their utmost objective is to extract from a large predictor space, a smaller subset of variables that portray strongest effects to yield an easily interpretable model and improve prediction accuracy (Tibshirani 1996). Penalization methods impose a penalty function of the regression coefficients on the likelihood function to shrink the coefficients towards zero, whereby those of insignificant variables are set to zero. Fan and Li (2001) posit that a good penalty function is one that provides estimators that simultaneously satisfy the mathematical conditions of sparsity, unbiasedness and continuity. Developed to overcome the flaws of the traditional lasso (i.e., least absolute shrinkage and selection operator), the adaptive lasso and the SCAD (i.e., smoothly clipped absolute deviation) penalty functions simultaneously satisfy these properties (Fan and Li 2001; Zou 2006; Zhang and Lu 2007). With proper choice of regularization parameters, these references have demonstrated that SCAD and adaptive lasso perform as well as if the subset of significant variables were known in advance. This is referred to as the oracle property.

Methodological work on penalized variable selection in survival analysis date back to nearly three decades, with the work of Tibshirani (1997), and Fan and Li (2002) being the pioneering studies. Tibshirani considered univariate survival data (e.g., time-to-death), and developed a lasso variable selection method via a partial likelihood constrained optimization for the Cox proportional hazards (PH) model. Fan and Li's approach is based on a penalized semiparametric gamma frailty model with SCAD penalty function.  An advantage is that formulation of the penalized likelihood is straightforward because the marginal likelihood of the model is available in a closed form where frailty is allowed to follow the gamma distribution. Log-normal distribution is commonly used in frailty models (e.g., see McGilchrist 1993). However, it does not provide a closed form expression for the marginal likelihood. Ha et al. (2014) used a hierarchical likelihood approach which simplifies the calculation of the marginal likelihood to provide a penalize likelihood. Xie et al. (2020) applied Gaussian quadrature methods to approximate the integral in the marginal likelihood. Despite the relatively long history of methodological work on variable selection in
survival analysis, work focusing on mixture cure models have not received adequate attention. Studies along this line stem from the work of Liu et al. (2012),  Scolas et al. (2016), and Masud et al. (2018). The work of Masud et al. considered penalized variable selection in both mixture and promotion time cure models. Recently, Beretta and Heuchenne (2021) proposed a SCAD penalized variable selection method for the semiparametric PH cure model due to Sy and Taylor (2000). The underlying computational procedure is implemented in the R package ``\texttt{penPHcure}". Instead of the usual Cox PH model, Parsa et al. (2024) utilized the accelerated failure time (AFT) model, and developed a penalized AFT mixture cure model based on the adaptive lasso penalty function. Fu et al. (2022) developed a fully parametric penalized mixture cure model which provides variable selection in both low- and high-dimensional setting. However, all these methods are customized to either univariate survival data (e.g., time-to-death) or interval-censored data, and so they are impractical for correlated survival data. Clearly, methodological work on variable selection for correlated survival data, and univariate survival data with cure fraction have been considered in separate studies in a disaggregated fashion. This leaves a vital literature gap concerning practical situations requiring the intersection of correlated time-to-event endpoints and cure fraction, which typically occur in longitudinal studies involving recurrent event outcomes. 

Motivated by this background, the aim of this work is to develop a multivariate penalized variable selection method for recurrent data with a cure fraction. To allow for cure fraction and intra-subject correlation, we develop the method within frailty mixture cure models (Tawiah et al. 2020a), leveraging several methods including the generalized linear mixed model (GLMM) methodology (McGilchrist 1994; McGilchrist and Yau, 1995), inference of mixture models (McLachlan and Peel 2000), and penalized likelihood techniques (Fan and Li, 2001; Wang and Leng 2007; Zou 2006). Accordingly, we establish a penalized likelihood formed on the basis of an extended best linear unbiased prediction (BLUP) procedure in GLMM, which circumvents the intractable integration associated with the underlying marginal likelihood. Estimation is facilitated by an expectation--maximization (EM) algorithm, which provides a penalized restricted maximum likelihood (REML) method for variable selection and estimation of variance components. We consider adaptive lasso and SCAD penalty functions in the penalized estimation method. The proposed method is illustrated using Monte Carlo simulation studies and a real data example. To our knowledge, Kızılaslan et al. (2025) is the only known work that considered penalized variable selection in frailty mixture cure models. They used the elastic-net penalty function for penalization, and allowed the frailty term to follow the gamma distribution. Nonetheless, the method was developed for univariate survival data, and so its validity in correlated survival data such as recurrent events data is not yet known.

\section{Data notation and models}\label{model}
\subsection{Frailty mixture cure models}
Suppose recurrent event data are available on $i=1,...,m$ patients (also known as subjects) with $ j=1,...,n_{i}$ defined as the number of recurrent events observed on the $i$th patient, where $n=\sum_{i=1}^{m}n_{i}$ is the total number of recurrent event observations. Denote by $a_{ij}$ and $c_{ij}$ the uncensored and the right censored gap time between any $j$
successive observations on the $i$th patient. In practice, we observe $t_{ij}=\mbox{min}\left\lbrace a_{ij}, c_{ij} \right\rbrace  $ the gap time between the $j$th and the $(j-1)$th recurrent event observations on the $i$th patient and $\delta_{ij}=1(a_{ij}\le c_{ij})$ the associated censoring indicator.  We note that $c_{ij}$ is conditionally independent of $a_{ij}$ given $x_{i}=(x_{i1},...,x_{id})^{T}\in \mathbb{R}^{d}$ and $z_{i}=(z_{i1},...,z_{ip})^{T}\in \mathbb{R}^{p}$, the observed $d$- and $p$-dimensional covariates characterizing the cure probability and the risk of experiencing recurrent events, respectively. Note that in some real world applications, $x_{i}$ and $z_{i}$ may overlap either partially or completely.  

McLachlan and Peel (2000) discuss variants of mixture models with many useful applications in survival analysis. The mixture cure model describes a marginal survival function for a population that consists of a subgroup of patients who are not susceptible to the event of interest (cured) and those who are at greater risk (uncured). Originally developed for univariate survival data (e.g., Sy and Taylor 2000), its multivariate generalization has been proposed for recurrent events data (Yu 2008; Rondeau et al. 2011; Tawiah et al. 2020a), with the marginal survival function written as  
\begin{equation}\label{mixture}
S\left( t_{ij}\right)=1-\pi_{j}\left(y_{ij};x_{i} \right) +\pi_{j}\left( y_{ij};x_{i}\right)S_{u}(t_{ij};z_{i}),  
\end{equation}   
where  $y_{ij}$ is a binary group indicator, with $y_{ij}=1$ denoting that subject $i$ will eventually experience the $j$th recurrent event (uncured), and $y_{ij}=0$ indicating that subject $i$ will never experience the $j$th event (i.e., cured patient). The component $\pi_{j}\left(. \right) =Pr\left(y_{ij}=1; x_{i}\right) $ known as the incidence is the conditional probability of experiencing the $j$th event after the $(j-1)$th event and $1-\pi_{j}\left(. \right) = Pr\left( y_{ij}=0; x_i\right)$  defines the conditional probability of cure after each event. The term $S_{u}(.)$ is the conditional survival function of the uncured patients (termed as latency). For the cured patients, the component conditional survival function is degenerate concentrated on 1.  The probability $\pi_{j}\left(. \right)$ can be  constrained to the first event per subject, neglecting subsequent events.  However, as we consider in (\ref{mixture}) incorporating all available event occurrences increases statistical power and provides a more general model that models the probability of cure beyond first or initial events. Thus, such a model allows the assessment of the chances of cure after each event which gives a broad picture of the outcome of interest. 

For the multivariate marginal survival function (\ref{mixture}), the event times (i.e., gap time) $t_{ij}$ within a patient are potentially correlated due to dependence of sequential events. Furthermore, the binary indicator $y_{ij}$ that defines the cure status may be correlated with the event times $t_{ij}$ because they both emanate from the same patient. Omitting the correlation may lead to incorrect statistical inference. Joint modeling has become a fundamental framework for modeling correlated outcomes, using random effects or a marginal modeling approach (Fitzmaurice and Laird 1995; He et al. 2015; Tawiah and Bondell 2023). For example, Tawiah and Bondell (2023) considered correlated binary and survival outcomes in clustered data, proposing a multilevel joint model with nested random effects. They shown that ignoring the correlation increases the bias of fixed effects, and variance component parameters, and produces coverage probabilities that do not maintain the nominal level. In frailty mixture cure models, Tawiah et al. (2020b) incorporated dependent censoring by way of modeling correlation between recurrent events and a terminal event. In their model, separate random effects were specified for the recurrent events and the terminal event, with the correlation modeled through a covariance matrix formed by the joint distribution of the random effects. Ng et al. (2023)
studied dependent censoring in joint models relaxing the cure fraction assumption. On the other hand, Liu et al. (2004) used shared frailty models, utilizing a single random effect term to model the correlation between recurrent events and the terminal event. Following their approach, we propose to use a single random effect that is shared by  $y_{ij}$ and $t_{ij}$  to model the correlation in (\ref{mixture}). The random effect can be regarded as an unobserved covariate (i.e., frailty) that induces the correlation (McGilchrist 1993). As opposed to the above work that utilized multiple random effects, the use of a single shared random effect reduces model complexity, and hence may enhance computational efficiency and minimize model non-identifiability issues.  Let $u_{i}$ denote the random effect of the $i$th patient and $u=(u_{1},...,u_{m}) $ an $m$-dimensional vector of $u_{i}$ assumed to follow the normal distribution $ N(0, D(\theta))$ with covariance matrix $D(\theta)$. Noting the conditional independence of $t_{ij}$ and $y_{ij}$ given $x_{i}$, and $z_{i}$, we propose to model jointly the incidence and the latency of (\ref{mixture}) via $u_{i}$, using a shared parameter logistic regression model and Cox PH frailty model, specified as follows  
\begin{equation}\label{latinc}
\begin{aligned}
\pi\left(y_{ij};x_{i} \right)&=\exp\left( \xi_{ij}\right)/\left[1 +\exp\left(\xi_{ij} \right)  \right]  ; \hspace{0.2cm} \xi_{ij}  =w^{T}_{i}\alpha+u_{i},\\
\lambda\left(t_{ij};x_{i} \right)&=\lambda_{u0}\left(t_{ij} \right)\exp\left( \eta_{ij} \right); \hspace{1.2cm} \eta_{ij}  =z^{T}_{i}\beta+u_{i},
\end{aligned}   
\end{equation} 
where $w_{i}=( 1 \hspace{0.2cm} x^{T}_{i})^{T} $, $\xi_{ij}$ and $\eta_{ij}$ are the linear predictors corresponding to the incidence and the latency. Also, $\alpha=(\alpha_{0},\alpha_{1},...,\alpha_{d})\in\mathbb{R}^{(d+1)}$ and  $\beta=(\beta_{1},...,\beta_{p})\in\mathbb{R}^{p}$ are the fixed effect parameter vectors corresponding to $\pi_{j}(.)$ and $t_{ij}$, respectively, and $\lambda_{u0}(.)$ is the baseline hazard function for the distribution of $t_{ij}$ which may be modeled parametrically or nonparametrically. Unification of models ($\ref{mixture}$) and ($\ref{latinc}$) refers to the frailty mixture cure model, which includes several models as special cases. The model can be referred to as a log-normal frailty mixture cure model since the exponential function of the random effect on the normal distribution is log-normal.

We remark that the covariance matrix $D(\theta)$ is general and may take several forms. We consider the form $D(\theta)=\theta I_{m}$, where $I_{m}$ is an $m \times m$ identity matrix, which yields a frailty term that is constant over time (McGilchrist, 1993). On the other hand, the model can be generalized, using a multivariate random effect, that is, replacing $u_{i}$ with $u_{ij}$. This allows each $j$th recurrent event observation on the $i$th patient to be associated with its own frailty (Tawiah et al. 2020a). On this note, $u_{ij}$ follows a multivariate normal distribution $N(0, \theta G(\rho))$, where the covariance matrix $D(\theta)=\theta G(\rho)$ is a $n \times n$ block diagonal matrix, parameterized by an autoregressive order 1 (AR(1)) process (see Appendix A). Note that the random effect (i.e., either $u_{i}$, or $u_{ij}$) is shared by $t_{ij}$ and $y_{ij}$, and the recurrent event observations within the $i$th patient, and so it summarizes the correlation from both sources. This reflects the uniqueness of our proposed model, and hence estimation algorithms and software packages for fitting existing frailty mixture cure models are not directly applicable. In what follows, we present an estimation and variable selection method (Section \ref{meth}) for the model. 

\section{Estimation and variable selection method}\label{meth}
\subsection{BLUP log-likelihood}  
We note that the binary cure status indicator $y_{ij}$ and the random effect $u_{i}$ are not observed. We let $y_{ij}=1$ when $\delta_{ij}=1$, but is unknown when $\delta_{ij}=0$.  Given the observed data $ O_{ij}=( t_{ij}, \delta_{ij},x_{i},z_{i} )$, define $ A_{ij}=( t_{ij}, \delta_{ij},x_{i},z_{i},y_{ij}, u_{i} )$ as the ``augmented'' data. The marginal likelihood of the ``augmented'' data associated with the frailty mixture cure model (\ref{mixture}-\ref{latinc}) takes the form  
\[
\mathcal{L}\left(\lambda_{u0},\alpha,\beta,u_{i} \right) =\int \prod_{i=1}^{m}\prod_{j=1}^{n_{i}} \mathcal{L}_{ij}^{A}\left(\lambda_{u0},\alpha,\beta, u_{i} \right)du_{i},
\]
where $\mathcal{L}_{ij}^{A}\left(\lambda_{u0},\alpha,\beta, u_{i} \right)=\exp( \ell_{ij}^{C}\left(\lambda_{u0},\alpha,\beta, u_{i} \right))g\left(u_{i},D(\theta) \right)  $ is the ``augmented'' data likelihood, with  $\exp( \ell_{ij}^{C}(\lambda_{u0},\alpha,\beta, u_{i} ))$ and $g\left(u_{i},D(\theta) \right)$ denoting the conditional likelihood of the ``augmented'' data and the density function of $u_{i}$, respectively. The marginal likelihood is analytically intractable due to integration over the random effect $u_i$ and its dependence on the partially missing cure status indicator $y_{ij}$. Assuming nonparametric form of $\lambda_{u0}(.)$ further complicates the marginal likelihood. To circumvent these complications we follow the GLMM methodology, utilizing the BLUP likelihood approach (McGilchrist 1994; McGilchrist and Yau 1995) which is essentially a penalized likelihood. Let $\ell_{1}$ denote the logarithm of the ``augmented'' data conditional likelihood and $\ell_{2}$ the logarithm of the density function of $u_{i}$. The BLUP likelihood is defined by $\ell(\Omega, \theta) = \ell_{1} + \ell_{2}$, where $\ell_2$ is the penalty term and $\Omega=(\alpha,\beta, u)^T$. 
From (\ref{mixture}) and (\ref{latinc}), $\ell_{1}$ is formed on the basis of the likelihood contribution of parameters in the logistic regression model and the Cox PH model with a shared random effect parameter. This can be expressed as
\begin{equation}
\ell_{1}=\log \left[\prod_{i=1}^{m} \prod_{j=1}^{n_{i}}\pi_{ij}^{y_{ij}}\left( 1-\pi_{ij}\right)^{1-y_{ij}}  \prod_{i=1}^{m} \prod_{j=1}^{n_{i}}  \left\lbrace \lambda_{u0}(t_{ij})\exp\left(\eta_{ij} \right)  \right\rbrace ^{\delta_{ij}y_{ij}}S_{u0}(t_{ij})^{y_{ij}\exp\left( \eta_{ij}\right) } \right], 
\end{equation} 
where $\pi_{ij}=\pi_{j}(x_{i})$, $\lambda_{u0}(t_{ij})=f(t_{ij})/S(t_{ij})$, and $S_{u0} \left(t_{ij} \right)=\exp \left\lbrace {- \Lambda_{u0}(t_{ij})} \right\rbrace $. Note that $\Lambda_{u0}\left( t_{ij} \right)$ is the cumulative baseline hazard function for the uncured subjects, given by $\Lambda_{0}\left( t_{ij} \right)=\int_{0}^{t_{ij}} \lambda_{u0}(s)ds $. In semiparametric PH models, the partial likelihood is frequently used, and it can be obtained through profile likelihood construction (Klein 1992) which cancels out $\lambda_{u0}(.)$. Accordingly, the BLUP log-likelihood $\ell(\Omega,\theta)$ simplifies to 
\begin{equation}\label{blup}
\ell\left(\Omega, \theta \right) =\sum_{r=1}^{n}\left\lbrace y_{r}\xi_{r}-\log\left[ 1 + \exp\left( \xi_{r}\right) \right]  \right\rbrace +
\sum_{r=1}^{q}\delta_{r}\left\lbrace \eta_{r}-\log \sum_{l\in R(r)}^{n} y_{l}\exp(\eta_{l})\right\rbrace +\ell_{2},
\end{equation}
where $y_{r}$, $\delta_r$,  $ \xi_{r}$ and $\eta_r$ are the vectors of the binary group indicator $y$, the censoring indicator $\delta$, and the linear predictors $\xi=W\alpha+Ru$, and $\eta=Z\beta+Ru$ reordered according to the ordering of the gap times in ascending order $t_{(1)}<...<t_{(q)}$. Also,  $\eta_{l}$ and $y_{l}$ are the respective values of $\eta$ and $y$ associated with the risk set $R(r)$. The log-likelihood $\ell_{2}$ is given by
\[
\ell_{2}= 
\begin{cases}
    1/2\left\lbrace m\log\left( 2\pi \theta \right) + \theta^{-1}u^{T} u  \right\rbrace, & \text{ } u \sim N(0, \theta I_{m})\\
   1/2 \left\lbrace n\log\left( 2\pi \theta\right) +\log  |G|+\theta^{-1}u^{T}G^{-1}u \right\rbrace,               & 
   \text{ }   u\sim N(0,\theta G(\rho)),
\end{cases}
\]  
where $|G|$ is the determinant of matrix $G$.

\subsection{Penalization and penalty functions}  
To provide penalized estimation for model characterized by  (\ref{mixture}) and (\ref{latinc}), consider the penalized log-likelihood of the form 
\begin{equation}\label{penblup}
\ell_{p} = \ell\left(\Omega, \theta \right) + n\left\lbrace \sum_{j=1}^{d}\varphi_{\kappa_1}\left(|\alpha_{j}| \right)  + \sum_{j=1}^{p}\varphi_{\kappa_2}\left(|\beta_{j}| \right)\right\rbrace ,  
\end{equation}
where $\ell\left(\Omega, \theta \right)$ is the BLUP log-likelihood (\ref{blup}), $\varphi_{\kappa_1}\left(|.| \right)$ and $\varphi_{\kappa_2}\left(|.| \right)$ are penalty functions corresponding to $\alpha$ and $\beta$ respectively, and $\kappa=\left(\kappa_{1}, \kappa_{2} \right) $ are the tuning parameters. The tuning parameters regulate the amount of shrinkage applied to $\alpha$ and $\beta$. We note that the intercept term $\alpha_{0}$ in the incidence submodel is not penalized. When random effect is released from the model, the penalized log-likelihood (\ref{penblup}) looses $\ell_{2}$ and is equivalent to that of a standard mixture cure models (Liu et al. 2012; Masud et al. 2018; Beretta and Heuchenne 2021). As recurrent events data are generally not independent, utilizing (\ref{penblup}) without random effect may lead to penalized estimators that are not asymptotically efficient as the oracle estimator.  

For the penalty functions we restrict our focus to adaptive lasso and SCAD due to their popularity and many desirable properties. Adaptive lasso is a convex optimization problem and it does not suffer local minimization issue (Zou, 2006). Conversely, SCAD may have several local maximizers. However, asymptotically the local maximizer of SCAD satisfies the oracle property and performs as well as an oracle maximum likelihood (ML) estimator (Fan and Li, 2001).  For $\alpha$ and $\beta$ the adaptive lasso penalty function is respectively given by $\sum_{j=1}^{d}\varphi_{\kappa_1}\left(|\alpha_{j}| \right)=\kappa_1\sum_{j=1}^{d}w_{1j}|\alpha_{j}|$ and $\sum_{j=1}^{p}\varphi_{\kappa_2}\left(|\beta_{j}| \right)=\kappa_2\sum_{j=1}^{p}w_{2j}|\beta_{j}|$, where $w_{1j}=(w_{1},...,w_{d})^T$ and $w_{2j}=(w_{1},...,w_{p})^T$ are data-dependent weight vectors. 
Given the weight vectors, adaptive lasso applies different amounts of shrinkage to the regression coefficients. It allows smaller coefficients to be severely penalized to shrink their estimates to zero, whereas larger coefficients receive smaller penalties to avoid unnecessary biases.  
We take $w_{1j}=1/|\tilde{\alpha}_{j}|$ and $w_{2j}=1/|\tilde{\beta}_{j}|$, where $\tilde{\alpha}=(\tilde{\alpha}_{0}, \tilde{\alpha}_{1}, ..., \tilde{\alpha}_{d})^T$  and $\tilde{\beta}=(\tilde{\beta}_{1}, ..., \tilde{\beta}_{p})^T$. We note that $(\tilde{\alpha}, \tilde{\beta} ) = \mbox{argmax} \ell\left(\Omega, \theta \right)$ is the global maximizer of the BLUP log-likelihood (\ref{blup}) which is elaborated in a later paragraph. If $\theta$ is found by REML then $(\tilde{\alpha}, \tilde{\beta} ) $ is the REML estimate, otherwise it is ML estimate if $\theta$ is estimated by ML. We focus on REML because its superiority to ML is widely documented, in terms of accuracy of empirical bias of variance components (e.g., see Tawiah et al., 2019). For SCAD the penalty function in terms of $\alpha$ takes the form    
\[
    \varphi_{\kappa_{1}}\left(|\alpha_{j}| \right) = 
\begin{cases}
    \kappa_{1}|\alpha_{j}|,& \text{ } |\alpha_{j}|\leq \kappa_{1},\\
 \frac{\left(a^2-1 \right)\kappa_{1}^2-\left(|\alpha_{j}| -a\kappa_{1}\right)^2  }{2\left( a-1\right) },  & \text{ } \kappa_1 < |\alpha_{j}| \leq a\kappa_{1} , \\
    \frac{\left( a +1\right)\kappa_{1}^{2} }{2},             & \text{ } |\alpha_{j}| > a\kappa_{1},
\end{cases}  
\]
where $a=3.7$ as suggested by Fan and Li (2001) to minimize the computational burden of SCAD optimization problem. We note that the SCAD penalty function for $\beta$ has the same form. Like adaptive lasso, the SCAD penalty allows larger coefficients to receive minimal amount of shrinkage in order to avoid excessive bias. In practice, one need to cleverly search for the best pair $\kappa=\left(\kappa_{1}, \kappa_{2} \right) $ over a two-dimensional grid for both SCAD and adaptive lasso. This is usually based on a data-dependent procedure using cross-validation, generalized cross-validation or a BIC criteria. We demonstrate our procedure for selecting the optimal pair $\kappa=\left(\kappa_{1}, \kappa_{2} \right) $ in Section \ref{tuning}.

\subsection{Penalized REML method}\label{penREML}
EM algorithms are the most commonly used methods for estimating the parameters of finite mixture models (McLachlan and Krishnan 2007). In mixture cure models, the linear predictors in the incidence, and the latency component are orthogonal. Hence, the underlying likelihood function naturally separates out into components of the logistic regression model, and the Cox PH model. This allows ease of programming the M-step separately using existing computational methods for the logistic regression model, and that of the PH model (Cai et al. 2012). When random effects are included, the M-step may be partitioned, utilizing methods for mixed effect logistic regression model, and those of a Cox frailty model (Tawiah et al. 2020a). Estimation in penalized mixture cure models follows suit, using computational methods for penalized logistic regression model, and penalized Cox PH model (Liu et al. 2012). However, in the frailty mixture cure model given in (\ref{mixture}) and (\ref{latinc}), orthogonality of the linear predictors $\eta_{ij}$ and $\xi_{ij}$ is ruled out by the shared random effect term. Therefore, the “augmented” data BLUP log-likelihood (\ref{blup}) does not separate out, and so the M-step cannot be implemented in the usual way. Additionally, calculations required in the E-step are intractable due to missingness of two variables, that is, the random effect $u_{i}$ and the binary group indicator $y_{ij}$. The M-step can be complicated in penalized mixture cure models,  particularly when using the SCAD penalty function because it is nonconcave and may not always guarantee a second order derivative. This can be remedied using a local quadratic approximation (LQA) of SCAD to obtain a modified Newton-Raphson algorithm (Fan and Li 2001) or perturbing SCAD slightly to make it differentiable (Hunter and Li 2005). 

To obtain penalized estimates, the M-step can be carried out through maximization of (\ref{penblup}). By this way, regularization (see Section~\ref{tuning}) enters the M-step, including variance component estimation which alternates with the E-step. The drawback of this approach is that it is computationally time consuming. The non-separability of the M-step adds additional computational cost since it involves iterative computation of large matrices. From here onwards, we assume the incompleteness of data is only due to the missingness of the binary group indicator variable $y_{ij}$. The random effect $u_{i}$, on the other hand, is conditionally fixed given the ``augmented" data BLUP log-likelihood (\ref{blup}) (McGilchrist 1994; McGilchrist and Yau 1995). The E-step calculates the conditional expectation of (\ref{blup}), which  simplifies to     
\begin{equation}\label{estep}
   g_{r}=E\left\lbrace y_{ij(r)}|\left( \Omega, S_{u0}(.)\right) \right\rbrace   = \delta_{ij(r)}+\frac{\left(1-\delta_{ij(r)} \right)\pi\left(x_{i(r)} \right)S_{u0}\left(t_{ij(r)} \right)^{exp\left(\eta_{ij(r)} \right) }   }{1-\pi\left(x_{i(r)} \right)+ \pi\left(x_{i(r)} \right)S_{u0}\left(t_{ij(r)} \right)^{exp\left(\eta_{ij(r)} \right) }},
  \end{equation}   
where $g_{r}$ is the posterior probability of being uncured sorted in the $r$th order according to $t_{(1)}<...<t_{(q)}$. The baseline survival function $S_{u0}(.)$ in (\ref{estep}) is estimated by a Breslow-type estimator (Sy and Taylor, 2000) integrated with the ETAIL completion method (Peng and Yu, 2021) given by
\begin{equation}
\hat{S}_{u0}\left(t_{ij(r)} \right)=
    \begin{cases}
       \exp \left\lbrace - \frac{\sum_{j:t_{(j)}\le t} d_{j(r)}}{\sum_{l \in R(t_{j})}g_{l}\exp\eta_{l}} \right\rbrace, & \text{ } t_{ij(r)}\le t_{h},\\
       \exp\left(- \hat{\psi} t_{ij(r)} \right), & \text{ } t_{ij(r)}>t_{h},
    \end{cases}
\end{equation}
where $t_{h}$ is the largest uncensored gap time and $\hat{\psi}=-\log \left\lbrace  S_{u0}\left( t_{h}\right)\right\rbrace /t_{h} $. The ETAIL method is driven by the exponential distribution, ensuring that $S_{u0}(.)$ smoothly goes to zero after $t_{h}$ to avoid non-identifiability. In principle, other tail completion methods, such as zero-tail, and WTAIL (based on Weibull distribution) could be used. The posterior probability $g_{r}$ updates $y_{r}$ in the ``augmented" data BLUP log-likelihood (\ref{blup}) and is maximized at the M-step using the Newton-Raphson algorithm
\begin{equation}\label{newton}
   \widetilde{\Omega}^{(k)}= \widetilde{\Omega}^{(k-1)} + \Sigma^{-1}\left(\frac{\partial \ell(\Omega, \theta)}{\partial \Omega} \right) ,
   \end{equation}
where $\widetilde{\Omega}^{(k)}$ is the vector of REML estimates of $\Omega$ at the $k$th iteration, $\Sigma^{-1}$ is the inverse of the matrix of the negative second derivatives  $\Sigma=-\partial^2 \ell(\Omega,\theta)/\partial\Omega\partial\Omega^2$ and the score vector $\partial \ell(\Omega, \theta)/\partial \Omega$ has the following simplification 
\[\frac{\partial \ell(\Omega,\theta)}{\partial \alpha}=W^{T}\frac{\partial \ell_{1}}{\partial \xi}; \hspace{0.2cm}
\frac{\partial \ell(\Omega,\theta)}{\partial\beta}=Z^{T}\frac{\partial \ell_{1}}{\partial \eta}; \hspace{0.2cm} 
\frac{\partial \ell(\Omega,\theta)}{\partial u}=R^T\left(\frac{\partial \ell_{1}}{\partial\xi} + \frac{\partial \ell_{1}}{\partial\eta}\right)-D(\theta)^{-1}u. 
\]
Note that $D(\theta)^{-1}=\theta^{-1}I_{m}$ when $u\sim N(0, \theta I_{m})$, and $D(\theta)^{-1}=\theta^{-1}G^{-1}$ when $u\sim N(0,\theta G(\rho))$. The derivation details of $\partial \ell_{1}/\partial\xi$, $\partial \ell_{1}/\partial\eta$, and $-\partial^2 \ell(\Omega,\theta)/\partial\Omega\partial\Omega^2$ are provided in Appendix B. Because (\ref{blup}) is non-separable the matrix $\Sigma$ is non-separable too. Therefore, $\Sigma$ can be very massive in dimension in applications with large sample size. This complicates its inversion involved in the Newton-Raphson iteration (\ref{newton}), and so leads to computational challenges in terms of time and space. As a remedy, we use the Cholesky decomposition of $\Sigma$ denoted by $\Sigma^{*}$ as an estimate for $\Sigma^{-1}$ as in Tawiah and Bondell (2023) which speeds computations considerably. For ease of presentation, we write $\Sigma$ as 
\[
\Sigma=\begin{bmatrix}
\zeta_{\alpha,\alpha} & \zeta_{\alpha,\beta} & \zeta_{\alpha,u}  \\
\zeta_{\beta,\alpha} & \zeta_{\beta,\beta} & \zeta_{\beta,u}  \\
\zeta_{u,\alpha} & \zeta_{u,\beta} & \zeta_{u,u}  \\
\end{bmatrix} \hspace{0.2cm} \mbox{and} \hspace{0.2cm} 
\Sigma^{*}=\Sigma^{-1}=\begin{bmatrix}
\Sigma^{*}_{\alpha,\alpha} & \Sigma^{*}_{\alpha,\beta} & \Sigma^{*}_{\alpha,u}  \\
\Sigma^{*}_{\beta,\alpha} & \Sigma^{*}_{\beta,\beta} & \Sigma^{*}_{\beta,u}  \\
\Sigma^{*}_{u,\alpha} & \Sigma^{*}_{u,\beta} & \Sigma^{*}_{u,u}  \\
\end{bmatrix},\] 
denoting some block matrices with partitions conformable to $\alpha|\beta|u$. Following Yau and Ng (2001), the estimate of the variance component $\theta$ is found by the REML estimating equation 
\begin{equation}\label{reml}
\tilde{\theta}=\frac{\left[\mbox{tr}\left( \Sigma^*_{u,u}\right) + u^{T}u \right] }{m},
\end{equation}
for $D(\theta)=\theta I_{m}$, where $\mbox{tr}$ denotes the trace of a matrix. Note that if we replace $\Sigma^*_{u,u}$ in (\ref{reml}) by $\zeta^*_{u,u}=\zeta^{-1}_{u,u}$ then $\theta$ is found by ML estimation (Tawiah et al. 2019) and our unfolding estimation procedure will turnout to be penalized ML (penML) rather than penalized REML (penREML). Although, we do not proceed with penalized ML it is interesting to investigate how it compares with penalized REML in terms of variable selection, which we show using simulation studies in Section \ref{sim}. For $D(\theta)=\theta G(\rho)$ the computation of $\theta$ and $\rho$ is based on REML detailed in Appendix A.

We note that $\Sigma$ is the asymptotic covariance matrix of the unpenalized parameter estimates, that is, the fixed effects $\tilde{\alpha}$, $\tilde{\beta}$, and the random effect $\tilde{u}$. The existence of a common joint covariance matrix for fixed effects, and random effect generalizes the least squares approximation results for fixed effect models (Wang and Leng 2007) to the setting of mixed effect models of which the model given in (\ref{mixture}) and (\ref{latinc}) is a special case.  Following Wang and Leng a Taylor series expansion at $\tilde{\Omega} = (\tilde{\alpha}, \tilde{\beta}, \tilde{u})$ provides the following approximation to the ``augmented" data BLUP log-likelihood (\ref{blup})   
\begin{equation}\label{taylor}
\ell\left( \Omega, \theta \right)  \approx \ell\left( \tilde{\Omega},\tilde{\theta} \right)  + \ell'\left(  \tilde{\Omega},\tilde{\theta}\right) ^T(\Omega -\tilde{\Omega} ) + \frac{1}{2}\left( \Omega -\tilde{\Omega} \right) ^T\ell''\left(  \tilde{\Omega},\tilde{\theta}\right) \left( \Omega -\tilde{\Omega} \right)  ,
\end{equation}
where $\ell'( .)$ and $\ell''( .)$ are the first- and second-order derivatives of the BLUP log-likelihood. As in Wang and Leng (2007) we take  $E\left\lbrace  \ell''(\tilde{\Omega},\tilde{\theta} )\right\rbrace  \sim \Sigma^{-1}$ and so $\tilde{\Sigma}^{-1}= \ell''(\tilde{\Omega},\tilde{\theta} )$ is a natural estimate for $\Sigma^{-1}$. Note that $\Sigma^{-1}$ is available at convergence of the EM algorithm in the form of $\tilde{\Sigma}^{*}$. Setting $\ell'( .)=0$ at $\tilde{\Omega}$ and omitting the constant $\ell( .)$ and the coefficient $1/2$, the approximation (\ref{taylor}) simplifies to the least squares approximation (LSA)  
\begin{equation}\label{lsa}
\ell(\Omega, \theta ) \approx \left( \Omega -\tilde{\Omega} \right) ^T\tilde{\Sigma}^*\left( \Omega -\tilde{\Omega} \right).
\end{equation} 
Replacing the BLUP log-likelihood in (\ref{penblup}) with the LSA (\ref{lsa}), we have  
\begin{equation}\label{apenblup}
Q(\Omega,\theta) = (\Omega -\tilde{\Omega} )^T\tilde{\Sigma}^*(\Omega -\tilde{\Omega} ) + n\left\lbrace \sum_{j=1}^{d}\varphi_{\kappa_1}\left(|\alpha_{j}| \right)  + \sum_{j=1}^{p}\varphi_{\kappa_2}\left(|\beta_{j}| \right)\right\rbrace  
\end{equation}
as a penalized least squares approximation of (\ref{penblup}). Given $\tilde{\Omega}$ and $\tilde{\theta}$ from (\ref{newton}) and (\ref{reml}) respectively, the approximate penalized BLUP log-likelihood (\ref{apenblup}) can be maximized, whereby the optimal pair  $\kappa=(\kappa_{1},\kappa_{2})$ is searched over a two dimensional grid via regularization (see Section \ref{tuning}). For easy of computation, (\ref{apenblup}) is partitioned to provide componentwise maximization without compromising the accuracy of parameter estimates. Let $\hat{\alpha}_{\kappa_1}=(\hat{\alpha}_{0},\hat{\alpha}_{\kappa_1,1},...,\hat{\alpha}_{\kappa_1,d})$ and
$\hat{\beta}_{\kappa_2}= (\hat{\beta}_{\kappa_2,1},...,\hat{\beta}_{\kappa_2,p})$ denote the penalized fixed effect estimates of $\alpha$ and $\beta$. Again, let $\hat{\Omega}_\kappa=(\hat{\alpha}_{\kappa_1}, \hat{\beta}_{\kappa_2}, \hat{u})$ denote a parameter vector for the penalized fixed effect estimates, and the random effect estimates. To obtain $\hat{\alpha}_{\kappa_1}$ we hold $\beta$ and $u$ constant, and then (\ref{apenblup}) is maximized replacing $\tilde{\Sigma}^*$ with $\tilde{\Sigma}^*_{\alpha,\alpha}$. The estimate $\hat{\beta}_{\kappa_2}$ is obtained in the same manner, holding $\alpha$ and $u$ constant and making use of $\tilde{\Sigma}^*_{\beta,\beta}$. Although interest lies in obtaining penalized fixed effect estimates $\hat{\alpha}_{\kappa_1}$ and $\hat{\beta}_{\kappa_2}$, the random effect estimate $\hat{u}$ can also be estimated through the maximizer of (\ref{apenblup}). This can be achieved by utilizing $\Sigma_{u,u}^*$,  holding $\alpha$ and $\beta$ constant and setting tuning parameters $\kappa_1$ and $\kappa_2$ to zero in the maximizer. In fact, $\hat{u}$ so obtained from the maximizer of (\ref{apenblup}) 
is very close to $\tilde{u}$ computed from (\ref{newton}). Therefore, updating the variance component $\theta$ in (\ref{reml}), by replacing $\tilde{u}$ with $\hat{u}$ provides little difference, but comes with extra computational cost and so should be avoided.  If for some reason, random effect inference is required it can be done using $\tilde{\Sigma}^*$ which is available at convergence of the EM algorithm. The procedure involved is already established (Tawiah et al. 2019). We note that the variance component estimator (\ref{reml}) neither depends on the penalty functions nor the estimators of penalized estimates. Hence, the estimate of $\theta$ remains the same when using different penalty functions. However, the penalized fixed effect estimators depend on the random effect $u$ and the variance parameter $\theta$ due to the sub-matrices of $\tilde{\Sigma}^*$ appearing in the maximizer of (\ref{apenblup}), and in tuning parameter selection (see Section \ref{tuning}). This ensures that the non-independence of the data (i.e., intra-subject correlation) is maintained and adjusted for in the penalized estimation.

\subsection{Regularization and standard error estimation}\label{tuning}
Regularization is a crucial part of penalized regression for determining the optimal values for the tuning parameters $\kappa=(\kappa_{1},\kappa_{2})$. The performance of penalized likelihood estimators depend on the choice of regularization parameter and the penalty function. In terms of adaptive lasso and SCAD, it has been established that their estimators can be efficient as the oracle when regularization is data-dependent and cleverly chosen (Fan and Li 2001; Zou 2006; Zhang and Lu 2007). The BIC tuning parameter selector identifies the true model consistently and overcomes the problems of overfitting associated with the AIC-type selector and the generalized cross-validation method (Wang, Li, and Tsai, 2007; Zhang, Li, and Tsai, 2010). For our proposed procedure, the BIC selector is given as 
\begin{equation}\label{bic}
BIC\left( {\kappa}\right) =\left(\hat{\Omega}_{\kappa} - \tilde{\Omega} \right)^{T} \tilde{\Sigma}^{*}_{q}\left(\hat{\Omega}_{\kappa} - \tilde{\Omega} \right)
+ \log\left(  n\right)  v_{\kappa}/n,
\end{equation}
where $v_{\kappa}$ is the number of non-zero coefficients in $\hat{\Omega}_{\kappa}$ and 
\[\tilde{\Sigma}^{*}_{q}=\begin{bmatrix}
\tilde{\Sigma}^{*}_{\alpha,\alpha} & \tilde{\Sigma}^{*}_{\alpha,\beta}   \\
\tilde{\Sigma}^{*}_{\beta,\alpha} & \tilde{\Sigma}^{*}_{\beta,\beta}   \\
\end{bmatrix}.\]  
Note that information for the random effect $u$ is preserved in (\ref{bic}) due to the information matrix $\tilde{\Sigma}^{*}_{q}$. The optimal pair $\kappa=(\kappa_{1},\kappa_{2})$ is selected automatically through an exhaustive search on a two-dimensional grid by minimizing the BIC criterion (\ref{bic}) and maximizing the approximate ``augmented" data penalized BLUP log-likelihood (\ref{apenblup}) simultaneously. 

For standard error estimation, Fan and Li (2001) used the conventional sandwich formula with LQA for the penalty functions to obtain an asymptotic covariance matrix for the penalized estimates.  
Following their approach, we define the covariance matrix of $\hat{\Omega}_{\kappa}$ as
\begin{equation}\label{sandw}
\widehat{\mbox{cov}}(\hat{\Omega}_{\kappa})=\left\lbrace Q''(\hat{\Omega}_{\kappa},\hat{\theta})+n\Psi_{\kappa}(\hat{\Omega}_{\kappa})\right\rbrace^{*}\widehat{\mbox{cov}}\left\lbrace  Q'(\hat{\Omega}_{\kappa},\hat{\theta})\right\rbrace \left\lbrace Q''(\hat{\Omega}_{\kappa},\hat{\theta})+n\Psi_{\kappa}(\hat{\Omega}_{\kappa})\right\rbrace^{*},
\end{equation} 
where $Q'(\hat{\Omega}_{\kappa},\hat{\theta})$ and $Q''(\hat{\Omega}_{\kappa},\hat{\theta})$ are the first and second derivatives of $Q(\Omega,\theta) $ at $\hat{\Omega}_{\kappa}$. To save space the details of (\ref{sandw}) are reported in Appendix C. 

In the LSA setting, Wang and Leng (2007) obtained a simple solution for the asymptotic covariance by inverting the submatrix of the observed information matrix corresponding to nonzero fixed effect estimates of a given LSA estimator. This approach intentionally ignores the standard errors for zero fixed effect estimates because they are expected to be estimated as zero. To automate the process of obtaining zero and nonzero standard errors, we use $\tilde{\Sigma}^{*}$ as an estimator for $Q''(\hat{\Omega}_{\kappa},\hat{\theta})$ in the LQA sandwich formula (\ref{sandw}), knowing that $\tilde{\Sigma}^{*}$ comes in place of matrix inversion for the purpose of improving computational efficiency.

\section{Simulation studies}\label{sim} 
We outline simulation studies to assess the finite sample performance of the proposed REML penalized method (Section \ref{penREML}) using adaptive lasso and SCAD penalty functions in the underlying penalized ``augmented" data BLUP likelihood (\ref{apenblup}) and the associated standard error formula (\ref{sandw}). For clarity of exposition, we refer to the method by the name of the penalty functions, i.e., adaptive lasso and SCAD. We use an oracle estimator to benchmark the performance of adaptive lasso and SCAD estimators. The oracle estimator is an unpenalized REML estimator for an ideal model containing mainly the significant predictors corresponding to the nonzero components of $\alpha$ and $\beta$. 

To allow comparison across different sample sizes, we consider three patient samples $m=200$, $m=600$ and $m=1000$. We generate recurrent event data characterized by cure fraction from the frailty mixture cure model (\ref{mixture}-\ref{latinc}) as follows. First, we assume a maximum follow-up of 2000 days and number of recurrent events per patient in the range of $n_{i}=1-5$ inclusively. We consider eight predictors $x_{i}=(x_{i1},...,x_{i8})^T$ shared by the incidence and the latency submodels, where $x_{i} $ is standard normal with $\phi=0.5$ pairwise correlation between its components. We use  $\alpha = (1.2, $-$0.8, 0.6, 0, 0, 0, 1.0, 0, 0)$ and $\beta = ($-$1.5, 0, 0, 1.5, 0, 0.9, 0, 0)$ as true values for the fixed effect parameters. This indicates that $(x_1, x_2, x_6)$ and $(x_1, x_4, x_6)$ are the sets of significant predictors in the incidence and the latency, respectively. Recurrent event gap times $t_{ij}$ are simulated from the PH model (\ref{latinc}) with $\lambda_{u0}(.)$ following Weibull distribution with scale parameter $\mu$ and shape parameter $\tau$. If a patient is cured, the gap time is taken to be infinitely large and subsequently censored at $c_{ij}$, where $c_{ij}$ is the censoring times generated from a uniform distribution U(0, 2000), allowing $25\%$ $(\mu=0.8$; $\tau=0.3)$ and  40\%  $(\mu=0.02; \tau=0.8)$ censoring rates. The binary group indicator $y_{ij}$ is generated from the Bernoulli distribution with the probability of being uncured obtained from the logistic regression model (\ref{latinc}). The random effect $u_{i}$ is generated from $ N(0,\theta I_{i})$ and is shared by the logistic and the PH model, with the true values of $\theta$ chosen to be $0.5, 1.5, 2.0$ to examine the performance of the method under different magnitudes of correlation. Like previous studies (e.g., Zou, 2006), we considered 100 replications of the simulated data.

Results are reported in Table~\ref{t:one} and Table~\ref{t:two}. In Table~\ref{t:one} the column labelled ``correct'' gives the average number of zero coefficients in $\alpha$ and $\beta$ that are correctly set to zero (5 is best). The column labelled ``incorrect'' provides the average number of nonzero coefficients in $\alpha$ and $\beta$ that are incorrectly estimated as zero (0 is best). The average mean squared error (MSE) is a measure for prediction accuracy  (Zhang and Lu, 2007), calculated from $(\hat{\Omega}_{\kappa}-\Omega)^T\Psi (\hat{\Omega}_{\kappa}-\Omega)$ averaged over 100 replications, where $\hat{\Omega}_{\kappa}=(\hat{\alpha}_{\kappa_1},\hat{\beta}_{\kappa_2})^T$, $\Omega=(\alpha,\beta)^T$ and $\Psi$ is the population covariance matrix of the predictors (Tibshirani, 1996). As expected the oracle estimator has the optimal values of the average number of correct zeros (5) and the average number of incorrect zeros (0), and it provides the smallest MSE values compared to SCAD, and adaptive lasso (Table~\ref{t:one}). However, it is seen that the results from SCAD and adaptive lasso estimators draw closer to those of the oracle estimator as sample size $m$ increases. SCAD slightly outperforms adaptive lasso in terms of identifying the true subset of zero effects (i.e., ``Correct'') for the incidence and the latency component. In terms of selecting the true subset of nonzero effects (i.e., ``Incorrect'') - both methods remain competitively good for the incidence, and the latency. The MSE of SCAD and adaptive lasso decrease towards zero as sample size $m$ increases. SCAD provides the smallest MSE values, compared with adaptive lasso.  As the random effect variance component $\theta$ increases, adaptive lasso and SCAD estimators remain relatively stable in obtaining the true subsets of zero effects and the non-zero effects, though their MSE values have moderate increment. Analogously, MSE for the oracle estimator increases slightly as $\theta$ increases. Unsurprisingly, this is because a higher variance in the random effect implies there is more variability in the data that is not explained by the fixed effects, hence making the fixed effect estimates less precise. Apparently, simulations conducted under 25\% and 40\% censoring rates produced slight differences in results, demonstrating robustness of the proposed methods under varied scenarios of censoring.  

\begin{table*}[tbp]
\centering
\caption{Simulation results for the proposed penalized REML method with adaptive lasso and SCAD penalty for assessing variable selection and prediction accuracy.}
\label{t:one}
\begin{threeparttable}
\resizebox{15cm}{!}{
\begin{tabular}{lclcccccccc}
\toprule
 & && & \multicolumn{3}{c}{Incidence submodel} &  & \multicolumn{3}{c}{Latency submodel}    \\ 
  \cline{5-7} \cline{9-11}
$(m, n_{i})$ & $\theta$  & Method &  & Correct & Incorrect & MSE &  & Correct & Incorrect & MSE  \\ \hline
25\% censoring & &  &  &  &   & &  & & &  \\
(200, 1--5) & 0.5 & Adaptive lasso &  & 4.35 & 0.15&  0.186&  & 4.91 & 0.02 &  0.071\\
 &  & SCAD & & 4.68 & 0.21 &  0.170 &   & 4.95 & 0.05  & 0.068 \\
  & & Oracle &  & 5 & 0 &  0.061 &  & 5 & 0 &  0.030 \\
  & 1.5 & Adaptive lasso &  & 4.09 & 0.28 &  0.274 &  & 4.81 & 0.01 &  0.132 \\
  &  & SCAD & & 4.40 & 0.31 &  0.241 &   & 4.98 & 0.06 &   0.102 \\
  & & Oracle &  & 5 & 0 &  0.080 &  & 5 & 0 &  0.063\\
  & 2.0 & Adaptive lasso &  & 4.14 & 0.27 &  0.270&  & 4.76 & 0.06 &  0.179\\
  &   & SCAD & & 4.36 & 0.33 &  0.259&   & 4.99 & 0.02 &  0.075 \\
  & & Oracle &  & 5 & 0 &  0.086 &  & 5 & 0&  0.062 \\ 
 (600, 1--5) & 0.5 & Adaptive lasso &  & 4.78 & 0.10 &  0.076 &  & 4.95 & 0.03 &  0.057 \\
  &  & SCAD & & 4.98 & 0.07 &  0.049 &   &  5 & 0.07&  0.053 \\
  & & Oracle &  &  5 & 0 &  0.029 &  & 5 & 0&  0.016 \\
  & 1.5 & Adaptive lasso &  & 4.79 & 0.12 &  0.110 &  & 4.97 & 0 & 0.055 \\
  &  & SCAD & & 5 & 0.16 &  0.079 &   & 5 & 0 &  0.025\\
  & & Oracle &  & 5 & 0 & 0.035 &  & 5 & 0 & 0.022 \\
  & 2.0 & Adaptive lasso &  & 4.75 & 0.08 &  0.095&  & 4.96 & 0 &  0.057 \\
  &  & SCAD & & 4.96 & 0.08 &  0.065 &   & 4.99 & 0 &  0.031 \\
  & & Oracle &  & 5 & 0 & 0.036 &  & 5 & 0 & 0.030 \\
(1000, 1--5)  & 1.5 & Adaptive lasso && 4.93 & 0.03 & 0.056 && 4.97 & 0& 0.052 \\
  & & SCAD && 5 & 0.03 & 0.033 && 5 & 0 & 0.025\\
  & & Oracle && 5 & 0 & 0.023 && 5 & 0 & 0.022 \\
40\% censoring  &&&&&&&&&&\\
(200, 1--5)  & 0.5 & Adaptive lasso&&  4.42 & 0.24 & 0.196 &  & 4.92 & 0.04 & 0.106 \\
  && SCAD && 4.76 & 0.24 & 0.192 && 5 & 0.10 & 0.094 \\
  && Oracle && 5 & 0 & 0.057 && 5& 0 & 0.040\\
  & 2.0 & Adaptive lasso && 4.12 & 0.16 & 0.346 && 4.70 & 0.08 & 0.165 \\
   && SCAD && 4.42 & 0.14 & 0.210 &  & 4.94 & 0.02 & 0.111 \\
   && Oracle && 5 & 0 & 0.071 && 5 & 0 & 0.101 \\
(600, 1-5)   & 0.5 & Adaptive lasso && 4.50 & 0.20 & 0.087 && 4.93 & 0.03 & 0.059 \\
   && SCAD &&  4.90 & 0.13 & 0.070 &  & 5 & 0.03 & 0.037 \\
   && Oracle && 5 & 0 & 0.020 && 5 & 0 & 0.013\\
(1000, 1-5)   & 1.5 & Adaptive lasso && 4.70 & 0 & 0.046 && 4.95 & 0 & 0.056 \\
   && SCAD &&  4.90 & 0 & 0.044 &  & 5 & 0 & 0.034 \\
   && Oracle && 5 & 0 & 0.014 && 5 & 0 & 0.010 \\
\hline
\end{tabular}
}
\end{threeparttable}
\end{table*}

\begin{figure}[!htb]
\centering
\includegraphics[width=0.7\linewidth]{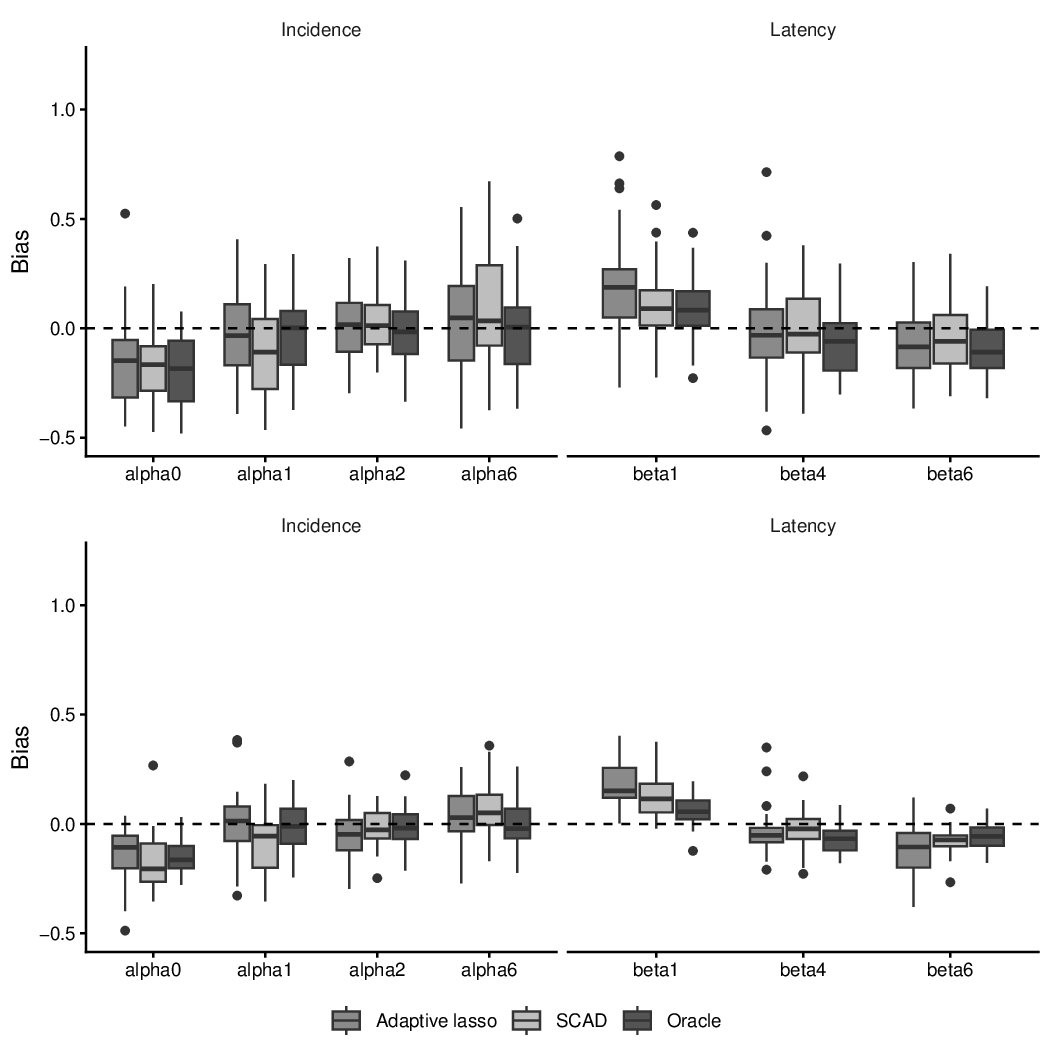}
\caption{Bias of nonzero fixed effects for the proposed penalized REML method with adaptive lasso and SCAD penalty, compared with the oracle estimator. Simulation is based on sample size $m=200$ (top panel), and $m=600$ (bottom panel) with 40\% censoring rate, and $\theta = 0.5$.}
\label{fig:fig1_bias}
\end{figure}

\begin{figure}[!htb]
\centering
\includegraphics[width=0.7\linewidth]{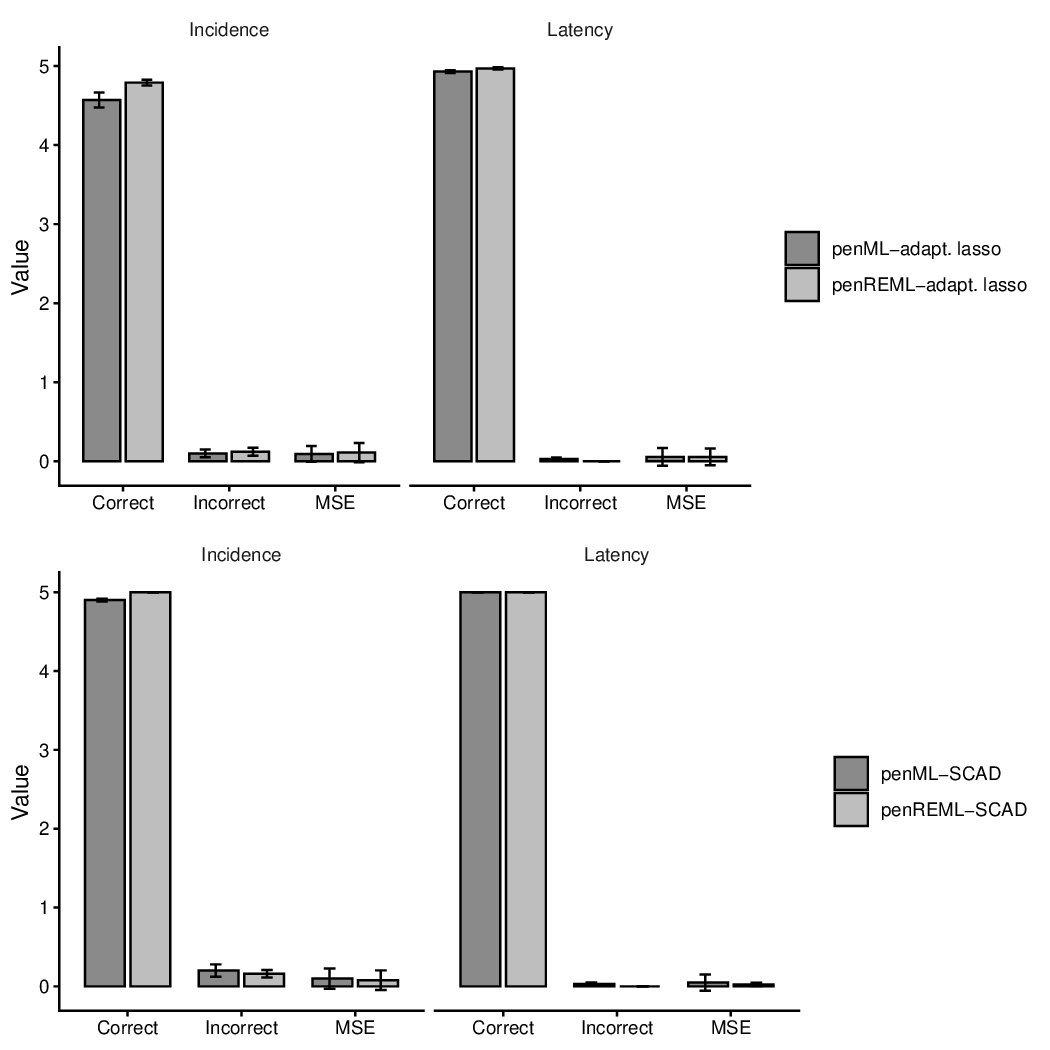}
\caption{Variable selection using penalized ML (penML), and penalized REML (penREML) with adaptive lasso (top panel) and SCAD (bottom panel) penalty functions. Simulation is based on sample size $m=600$, 25\% censoring and $\theta=1.5$.}
\label{fig:fig2_alasso_scad}
\end{figure}

Table~\ref{t:two} provides the average estimates of the nonzero coefficients of $\alpha$ and $\beta$, average of their asymptotic standard error (ASE) estimates and the empirical standard error (ESE) estimates. The ASE estimates for adaptive lasso and SCAD are obtained from the LQA sandwich formula (\ref{sandw}). Those of the oracle estimator are obtained from the information matrix of an unpenalized EM-based REML estimation of the frailty cure model containing only the significant variables. The ESE estimates are calculated to be the standard deviation of the parameters estimates. The average estimates of the nonzero coefficients for SCAD and the adaptive lasso are close to their true values and comparable to those of the oracle estimator for both low (25\%) and moderate (40\%) levels of censoring. As seen in Figure~\ref{fig:fig1_bias} the median bias for adaptive lasso, SCAD and the oracle estimator clearly overlap, showing no significant differences in the magnitude of the bias produced by these methods. Moreover, the bias of the three methods decreases towards zero as sample size $m$ increases.   
In terms of standard errors, the estimates of ASE and ESE from the oracle estimator have acceptable agreement. The ASE from the penalized estimators, particularly those of SCAD are close to the ASE estimates of the oracle estimator. For SCAD and adaptive lasso, the ESE estimates are slightly inflated in some simulation scenarios, when compared with their corresponding ASE estimates. This observation may be due to increased collinearity produced by insignificant variables that remain when the penalized methods overfit the model. As observed, the discrepancies between ASE and ESE are well pronounced where sample $m$ is small. Nevertheless, due to the robustness of the LQA sandwich formula (\ref{sandw}) the ASE estimates of SCAD, and adaptive lasso are unaffected by the increased collinearity, and they are as good as the ASEs produced by the oracle estimator. For the three methods, the mean estimates of the variance component parameter $\theta$  increase toward the true values as sample size $m$ increases. As discussed in Section~\ref{penREML}, $\theta$ has the same estimator under different penalty functions. Therefore, the mean estimates for $\theta$ are the same under adaptive lasso, and SCAD (Table~\ref{t:two}).  In Figure~\ref{fig:fig2_alasso_scad} we compare the performance of penalized REML estimators to that of penalized ML estimators in terms of variable selection. We use adaptive lasso and SCAD penalty functions in both estimators. From the figure it is clear that the REML penalized methods slightly outperform the ML counterparts. 

\begin{landscape}
\begin{table*}[!htb]
\caption{Simulation results for the nonzero components of $\alpha$ and $\beta$ from the proposed penalized REML method with adaptive lasso, and SCAD penalty using the constant frailty mixture cure model.}
\label{t:two}
\begin{threeparttable}
\resizebox{23.2cm}{!}{
\begin{tabular}{lcclccccccccccccccccccccc}
\toprule
 & & & & & \multicolumn{12}{c}{Incidence submodel}  & & & \multicolumn{5}{c}{Latency submodel}    \\  
  \cline{6-16} \cline{18-25}
 &  & & &  & \multicolumn{2}{c}{$\alpha_{0}=$1.2}  &  & \multicolumn{2}{c}{$\alpha_{1}=$ $-$0.8} & & \multicolumn{2}{c}{$\alpha_{2}=$0.6} &  & \multicolumn{2}{c}{$\alpha_{6}=$1.0} &  & \multicolumn{2}{c}{$\beta_{1}=$ $-$1.5} &&\multicolumn{2}{c}{$\beta_{4}=$1.5}&& \multicolumn{2}{c}{$\beta_{6}=$0.9}  \\ 
 \cline{6-7} \cline{9-10} \cline{12-13} \cline{15-16} \cline{18-19} \cline{21-22}  \cline{24-25}  
$(m, n_{i})$ & $\theta$ & Method &  & Mean $\theta$ & Mean  & ASE (ESE) & & Mean   & ASE (ESE) & & Mean & ASE (ESE) & & Mean & ASE (ESE) & & Mean & ASE (ESE) & & Mean & ASE (ESE) && Mean & ASE (ESE) \\ \hline
25\% censoring &&&&&&&&&&&&&&&&&&&&&&&&\\
(200, 1--5)  & 0.5  & Adaptive lasso & & 0.463 & 1.051 & 0.125 (0.239) &  & $-$0.738& 0.082 (0.295)  &  & 0.468& 0.103 (0.244) & & 1.045 & 0.160 (0.093) &&$-$1.361& 0.099 (0.182) && 1.459& 0.113 (0.180) && 0.845& 0.103 (0.212) \\
 &  & SCAD & & 0.463 & 1.080 & 0.125 (0.213) &  & $-$0.853 & 0.159 (0.223)  &  & 0.495 & 0.119 (0.298) & & 1.103 & 0.170 (0.236) && $-$1.429 & 0.115 (0.177) && 1.499 & 0.113 (0.166) && 0.843 & 0.104 (0.239) \\
  &  & Oracle & & 0.482 & 1.047 & 0.123 (0.155) &  & $-$0.877 & 0.163 (0.168) &  & 0.595 & 0.153 (0.164)& & 1.040 & 0.169 (0.193) && $-$1.443 & 0.115 (0.118) &  & 1.410 & 0.111 (0.117) && 0.854 & 0.108 (0.115)  \\
  & 1.5 & Adaptive lasso & & 1.277& 1.087 & 0.145 (0.261) & & $-$0.723& 0.087 (0.292)  &  & 0.420  & 0.096 (0.219) & & 0.974& 0.172 (0.286) && $-$1.320& 0.104 (0.191) && 1.444 & 0.138 (0.247) && 0.801& 0.127 (0.241) \\
  &  & SCAD & & 1.277  & 1.106 & 0.144 (0.230) &  & $-$0.781 & 0.164 (0.268) &  & 0.467 & 0.116 (0.238) & & 1.023 & 0.184 (0.252)&& $-$1.410 & 0.142 (0.189) &  & 1.485 & 0.138 (0.226) & & 0.823 & 0.131 (0.263)\\
 &  & Oracle & & 1.290 & 1.080 & 0.143 (0.186) &  & $-$0.826 & 0.183 (0.202)  &  & 0.556 & 0.161 (0.179)&  & 0.965 & 0.187 (0.204) && $-$1.420 & 0.146 (0.159) && 1.420 & 0.134 (0.156) & & 0.828& 0.140 (0.164)\\
  & 2.0 & Adaptive lasso & & 1.571 & 1.129 & 0.155 (0.254) &  & $-$0.703&  0.086 (0.299) &  & 0.443& 0.100 (0.232) & & 0.952 & 0.178 (0.281) && $-$1.303 & 0.102 (0.211) & & 1.425& 0.146 (0.279) && 0.761 & 0.129 (0.286)\\
  &  & SCAD &  & 1.571 & 1.103 & 0.152(0.224) &  & $-$0.764 & 0.169 (0.206)  &  & 0.465 & 0.123 (0.215) & & 0.947 & 0.186 (0.264) && $-$1.428 & 0.156 (0.161) && 1.476 & 0.147 (0.208) && 0.835 & 0.147 (0.193)\\
   &  & Oracle & & 1.576 & 1.080 & 0.150 (0.179) &  & $-$0.772 & 0.189 (0.189)  &  &  0.548 & 0.165 (0.184) & & 0.934 & 0.190 (0.212) && $-$1.400 & 0.156 (0.145) && 1.380 & 0.141 (0.159) && 0.812 & 0.149 (0.146) \\ 
(600, 1--5)  & 0.5 & Adaptive lasso & & 0.481& 1.140 & 0.073 (0.169) &  & $-$0.759 & 0.058 (0.155)  &  & 0.495 & 0.065 (0.114) & & 1.010 & 0.090 (0.184) && $-$1.370 & 0.058 (0.114) && 1.500 & 0.064 (0.123) && 0.818 & 0.058 (0.196) \\
  &  & SCAD & & 0.481 & 1.130 & 0.072 (0.124) &  & $-$0.835 &  0.092 (0.127) &  & 0.554 & 0.078 (0.183) & & 1.040 & 0.096 (0.151) && $-$1.420 & 0.065 (0.126) && 1.520 & 0.064 (0.109) && 0.818 & 0.057 (0.139) \\
  &  & Oracle & & 0.488 & 1.090 & 0.071 (0.119) &  & $-$0.835 & 0.093 (0.118)  &  & 0.566 & 0.086 (0.101)& & 1.020 & 0.097 (0.140)&& $-$1.440 & 0.066 (0.075)&& 1.430 & 0.064 (0.091)&& 0.867 & 0.062 (0.072)\\
   & 1.5 & Adaptive lasso & & 1.372 & 1.170& 0.084 (0.163) &  & $-$0.673 & 0.056 (0.113)  &  & 0.495 & 0.069 (0.111) & & 0.918 & 0.097 (0.182) && $-$1.330& 0.066 (0.115) && 1.440& 0.077 (0.130) && 0.784& 0.074 (0.133) \\
   &  & SCAD & & 1.372 & 1.150 & 0.082 (0.139) &  & $-$0.748 & 0.097 (0.179)  &  & 0.515 & 0.078 (0.135) & & 0.969 & 0.104 (0.141) && $-$1.410 & 0.082 (0.099) && 1.460 & 0.076 (0.102) && 0.854 & 0.079 (0.101) \\
   &  & Oracle & & 1.396 & 1.109 & 0.081 (0.104) &  & $-$0.757 &  0.103 (0.128) &  & 0.558 & 0.091 (0.107)& & 0.924 & 0.104 (0.137)&& $-$1.397 & 0.082 (0.089) && 1.401  & 0.075 (0.077)&& 0.818 & 0.079 (0.072)\\
  & 2.0 & Adaptive lasso & & 1.702 & 1.150 & 0.089 (0.171) &  & $-$0.660&  0.058 (0.187) &  & 0.501 & 0.072 (0.194) & & 0.897 & 0.100 (0.188) && $-$1.320 & 0.067 (0.119) && 1.450 & 0.081 (0.133) && 0.781 & 0.079 (0.136)\\
  &  & SCAD & & 1.702 & 1.160 & 0.087 (0.126) &  & $-$0.731 &  0.105 (0.159) &  & 0.534 & 0.086 (0.192) & & 0.943 & 0.109 (0.143)&& $-$1.390 & 0.088 (0.105)&& 1.450 & 0.080 (0.108)&& 0.830& 0.085 (0.110)\\
  &  & Oracle & & 1.716 & 1.150 & 0.087 (0.108) &  & $-$0.750 &  0.108 (0.129) &  & 0.558 & 0.093 (0.106)& & 0.916 & 0.111 (0.132)&& $-$1.400 & 0.088 (0.088)&& 1.400 & 0.080 (0.086)&& 0.825 & 0.086 (0.103)\\
(1000, 1-5)  & 1.5 & Adaptive lasso & & 1.453 & 1.150 & 0.065 (0.095)&& $-$0.628 & 0.047 (0.139)&& 0.515 & 0.057 (0.132)&& 0.888 & 0.075 (0.150)&& $-$1.300 & 0.052 (0.109)&& 1.480 & 0.059 (0.101)&& 0.767 & 0.057 (0.134)\\
  && SCAD & & 1.453 & 1.160 & 0.064 (0.115) && $-$0.710 & 0.080 (0.121) && 0.537 & 0.069 (0.134)&& 0.928 & 0.081 (0.095)&  & $-$1.380 & 0.063 (0.095) & & 1.460 & 0.058 (0.108)&& 0.838 & 0.061 (0.071)\\
  && Oracle && 1.468 & 1.120 & 0.064 (0.068)&& $-$0.716 & 0.080 (0.093)&& 0.556 & 0.071 (0.083)&& 0.901 & 0.081 (0.087)&& $-$1.390 & 0.063 (0.079)&& 1.410 & 0.058 (0.067)&& 0.826 & 0.061 (0.065)\\[6pt]
40\% censoring &&&&&&&&&&&&&&&&&&&&&&&\\
(200, 1-5) & 0.5 & Adaptive lasso && 0.444& 0.994 &  0.125 (0.236)&& $-$0.823 & 0.083 (0.195)&&  0.466 & 0.095 (0.299)&& 1.014 & 0.158 (0.260)&& $-$1.315 &  0.099 (0.193) && 1.474 & 0.119 (0.214)&& 0.778 & 0.100 (0.232)\\
 && SCAD && 0.444& 0.961 & 0.123 (0.232) && $-$0.908 & 0.154 (0.201) && 0.485 & 0.115 (0.304) && 1.080 & 0.166 (0.252) && $-$1.400 & 0.118 (0.151) && 1.503 & 0.120 (0.161) && 0.769 & 0.099 (0.293) \\
 && Oracle && 0.461 & 1.070 & 0.125 (0.164) && $-$0.837 & 0.165 (0.152) && 0.553 & 0.150 (0.180) && 1.050 & 0.170 (0.152) && $-$1.350 & 0.115 (0.138) && 1.390 & 0.113 (0.170) && 0.834 & 0.109 (0.096) \\
  & 2.0 & Adaptive lasso && 1.498 & 1.033 &  0.151 (0.230)&& $-$0.761 & 0.089 (0.289)&&  0.432 & 0.107 (0.247)&& 0.817 & 0.165 (0.262)&& $-$1.300 &  0.103 (0.224) && 1.384 & 0.150 (0.224)&& 0.729 & 0.121 (0.314)\\
 && SCAD && 1.498 & 1.037 & 0.151 (0.163) && $-$0.803 & 0.187 (0.255) && 0.508 & 0.147 (0.256) && 0.887 & 0.192 (0.232) && $-$1.391 & 0.157 (0.204) && 1.424 & 0.149 (0.206) && 0.808 & 0.145 (0.254) \\
 && Oracle && 1.507 & 0.982 & 0.150 (0.150) && $-$0.784 & 0.190 (0.211) && 0.557 & 0.167 (0.151) && 0.915 & 0.190 (0.155) && $-$1.360 & 0.157 (0.194) && 1.270 & 0.142 (0.173) && 0.810 & 0.151 (0.160) \\
 (600, 1-5) & 0.5 & Adaptive lasso && 0.464 & 1.055 &  0.072 (0.135)&& $-$0.805 & 0.061 (0.168)&&  0.449 & 0.058 (0.253)&& 1.023 & 0.092 (0.146)&& $-$1.317 &  0.061 (0.103) && 1.461 & 0.065 (0.111)&& 0.759 & 0.057 (0.189)\\
 && SCAD && 0.464 & 1.031 & 0.072 (0.129) && $-$0.887 & 0.093 (0.138) && 0.501 & 0.075 (0.216) && 1.059 & 0.096 (0.126) && $-$1.377 & 0.066 (0.098) && 1.481 & 0.065 (0.094) && 0.794 & 0.059 (0.162) \\
 && Oracle && 0.470 & 1.050 & 0.072 (0.080) && $-$0.816 & 0.093 (0.114) && 0.587 & 0.088 (0.144) && 0.997 & 0.097 (0.107) && $-$1.440 & 0.066 (0.120) && 1.440 & 0.064 (0.085) && 0.847 & 0.061 (0.101) \\
 (1000, 1-5) & 1.5 & Adaptive lasso && 1.447 & 1.054 &  0.063 (0.162)&& $-$0.711 & 0.048 (0.121)&&  0.496 & 0.057 (0.091)&& 0.880 & 0.073 (0.125)&& $-$1.288 &  0.054 (0.085) && 1.435 & 0.059 (0.135)&& 0.742 & 0.055 (0.134)\\
 && SCAD && 1.447 & 1.073 & 0.062 (0.137) && $-$0.755 & 0.079 (0.089) && 0.493 & 0.063 (0.180) && 0.921 & 0.079 (0.105) && $-$1.331 & 0.062 (0.150) && 1.475 & 0.060 (0.086) && 0.719 & 0.053 (0.156) \\
 && Oracle && 1.455 & 1.080 & 0.055 (0.047) && $-$0.771 & 0.070 (0.087) && 0.590 & 0.068 (0.083) && 0.939 & 0.073 (0.085) && $-$1.420 & 0.050 (0.072) && 1.460 & 0.050 (0.055) && 0.847 & 0.047 (0.051) \\ 
\hline
\end{tabular}
}
\end{threeparttable}
\end{table*}
\end{landscape}

\section{Application to breast cancer gene expression data}\label{appl}
\begin{figure}[!htb]
\centering
\includegraphics[width=0.6\linewidth]{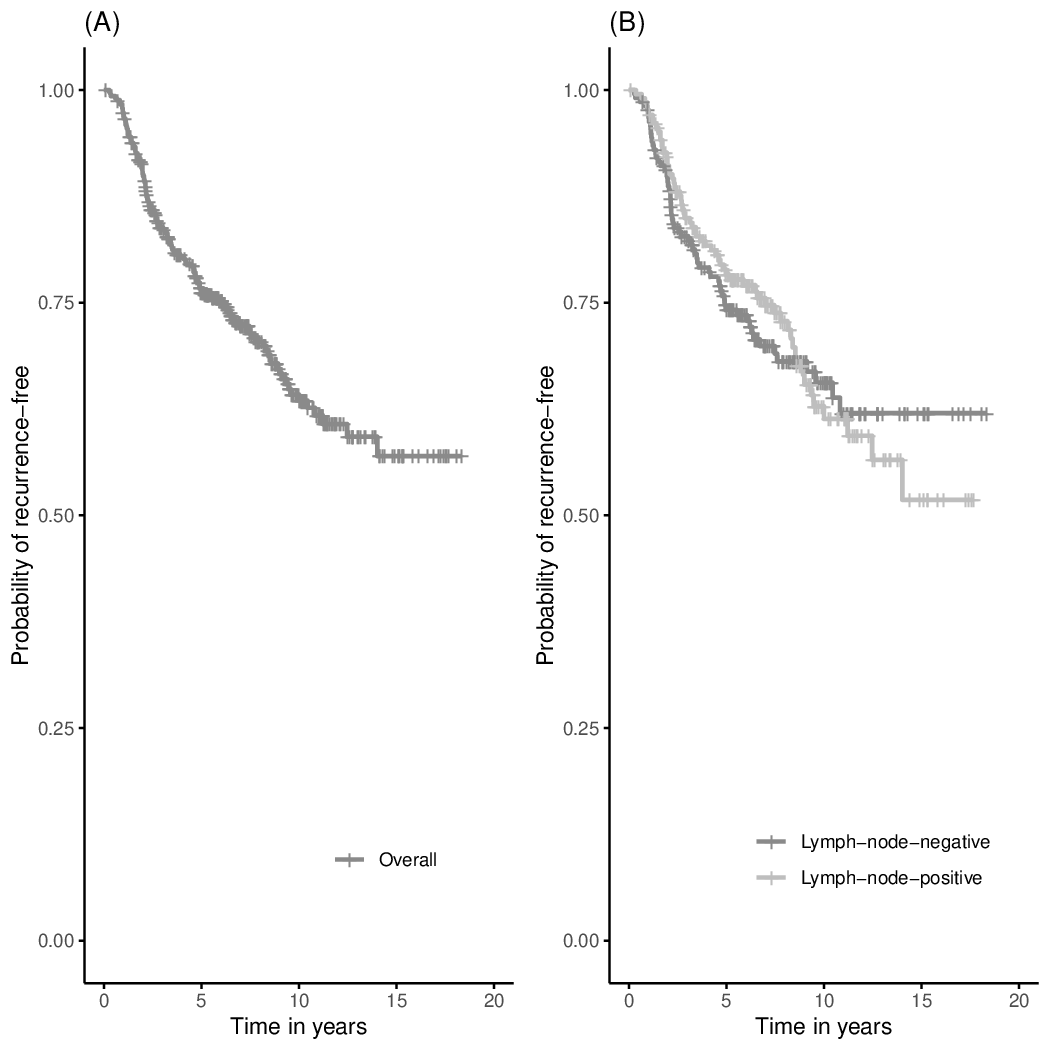}
\caption{Kaplan-Meier curve for the distribution of recurrence times of breast cancer (A) for the overall sample, and (B) stratified by lymph-node-disease status.}
\label{fig:fig3_KM}
\end{figure}

To illustrate the practical utility of the proposed methods we use a breast cancer gene expression data set described in Van De Vijver et al. (2002). The data set includes a series of 295 consecutive women selected from the fresh-frozen-tissue bank of the Netherlands Cancer Institute. In this study, we consider 294 samples due to one tissue having many missing gene expressions. Of 5153 genes available in this sample, 70 were previously classified as providing either good or poor prognostic value (Van't Veer et al., 2002). Due to the complexity of our model, and the relatively smaller sample size of the data, we consider 15 genes out of the 70 as potentially important predictors together with six clinical variables. The 15 genes were selected randomly. The clinical variables include surgery (0= breast-conserving therapy; 1=mastectomy), chemotherapy (0=no; 1=yes), hormonal therapy (0=no; 1=yes), lymph-node status (0=lymph-node-negative disease; 1=lymph-node-positive disease), tumour diameter (mean: 23.0mm ; standard deviation: 8.8mm) and age at diagnosis (mean: 43.5 years; standard deviation: 5.7 years). The data set provides failure time for three post-cancer surgery events, viz, metastasis, regional and local recurrence. Of the overall sample, 113 (38.4\%) experienced one to three post-cancer surgery events. Regarding these as recurrent events, from Figure~\ref{fig:fig3_KM}(A) the tail of the Kaplan-Meier curve for the distribution of the recurrence times levels off above 0.5 recurrence-free probability after 13 years of follow-up. Thus, it appears that no patient experienced any of the events beyond 13 years. Considering this information as the existence of potentially cure patients, we apply our proposed cure modeling-based penalized variable selection method to the data. We assume a constant frailty  $u\sim N(0, \theta I_{m}) $. Hence, the underlying model is a mixture cure model with constant frailty. Our aim is to identify the relevant variables that characterize recurrence risk and cure probability. All continuous covariates (gene expression, tumour diameter, age) are standardized. As noted in Section \ref{meth} we do not penalize the intercept term.

Results from an unpenalized frailty cure model and our proposed penalized model selection method with adaptive lasso and SCAD penalties are presented in Table~\ref{t:three}. The results are given in terms of fixed effect coefficients and their standard errors. In the case of the unpenalized model variable selection is based on $p$-values using a 5\% level of significance. The optimal values of the tuning parameters for adaptive lasso and SCAD are $\kappa =(0.003, 0.027)$ and $\kappa = (0.114, 0.021)$, respectively. As expected, the standard errors of the penalized models are smaller than those of the unpenalized model. For the clinical variables in the incidence part, the unpenalized model selects lymph-node-positive as the only significant variable associated with cure probability. Out of the six clinical variables adaptive lasso selects four (lymph-node-positive, chemotherapy, hormonal therapy, age) and SCAD retains five (lymph-node-positive, mastectomy, chemotherapy, hormonal therapy, tumour diameter). SCAD provides a small effect for mastectomy. For all the three methods lymph-node-positive has the strongest effect with positive coefficient estimate. Of the 15 gene expression variables, the unpenalized model selects six (AB037863, AF055033, NM016359, NM002019, NM014889, NM004702) for the incidence part. Adaptive lasso selects 11, namely AB037863, AF055033, AK000745, Contig46218RC, Contig55377RC, Contig63102RC, NM016359, NM002019, NM006101, NM014889, and NM004702. SCAD retains 10 (AA555029RC, AB037863, AF055033, AF257175, Contig46218RC, Contig55377RC, Contig32185RC, NM016359, NM014889, NM004702). Notably, SCAD excludes NM002019 which is identified as a significant predictor by the unpenalized model and adaptive lasso.  In the latency part, the unpenalized model identifies age and tumour diameter as the significant clinical predictors for recurrence risk of breast cancer in the uncured patients. Adaptive lasso retains five (lymph-node-positive, chemotherapy, hormonal therapy, age, tumour diameter), while SCAD selects all six clinical variables. The unpenalized model selects seven gene expression variables (AB037863, AK000745, AF257175, Contig32185RC, NM016359, NM002019, NM014889) for the latency part. Adaptive lasso identifies 12 (AA555029RC, AB037863, AK000745, AF257175, Contig55377RC, Contig32185RC, Contig63102RC, NM016359, NM002019, NM006101, NM014889, NM004702). SCAD selects 10, comprising the gene expressions identified by adaptive lasso with exception of AA555029RC and NM004702. Both AA555029RC and NM004702 identified by adaptive lasso have very small effect.

\begin{table*}[!htb]
\centering
\caption{Unpenalized REML method and the proposed penalized REML method with adaptive lasso and SCAD penalty applied to the breast cancer data. Underlying model is the constant frailty mixture cure model.}
\label{t:three}
\begin{threeparttable}
\resizebox{13.5cm}{!}{
\begin{tabular}{lccccccc}
\toprule
&  & \multicolumn{2}{c}{REML}  & \multicolumn{2}{c}{Adaptive lasso} & \multicolumn{2}{c}{SCAD}    \\ 
  \cline{3-4} \cline{6-6} \cline{8-8}
 Variable  & & Estimate (ASE)  & $p$-value &  & Estimate (ASE) &  & Estimate (ASE)  \\ \hline
\textbf{Incidence} & &  &  &  &   & &    \\
Intercept & & 0.011 (0.198) & 0.954 &  &  0.096 (0.023) & &  $-$0.020 (0.002)  \\
Lymph-node-positive & & 0.973 (0.353)& 0.006  &  &  1.080 (0.219) & &  0.994 (0.302) \\
Mastectomy & & $-$0.048 (0.247) & 0.845 &  &  0 (0) & &  0.004 (0.001)  \\
Chemotherapy & & $-$0.579 (0.355) & 0.102 &  &  $-$0.510 (0.179) & &  $-$0.584 (0.342)  \\
Hormonal therapy & & $-$0.051 (0.394) & 0.897 &  & 0.003 (0.001)  & &  0.017 (0.008)  \\
Age & & $-$0.218 (0.123) & 0.077 &  & $-$0.094 (0.058)  & & 0 (0)  \\
Tumour diameter & & $-$0.011 (0.127) & 0.931 &  & 0 (0)  & &  0.012 (0.003)  \\
AA555029RC & & 0.070 (0.138) & 0.610 &  &  0 (0) & & 0.027 (0.008)  \\
AB037863 & & $-$0.387 (0.135) & 0.004 &  &  $-$0.272 (0.101) & &  $-$0.431 (0.124)  \\
AF055033 & & 0.594 (0.143) & $<$0.001  &  & 0.589 (0.131)  & &  0.379 (0.114)  \\
AK000745 & & 0.145 (0.153) & 0.344 &  & 0.001 (0.001)  & &  0 (0)  \\
AF257175 & & 0.036 (0.137) & 0.793 &  &  0 (0) & &  0.001 (0)  \\
Contig46218RC & & 0.134 (0.188) & 0.477 &  & $-$0.001 (0.001)  & &  0.021 (0.006)  \\
Contig55377RC & & 0.295 (0.157) & 0.060 &  & 0.310 (0.114)  & & 0.223 (0.061)   \\
Contig32185RC & & 0.084 (0.156) & 0.587 &  &  0 (0) & &  $-$0.002 (0)  \\
Contig51464RC & & 0.009 (0.139) & 0.948 &  & 0 (0)  & & 0 (0)   \\
Contig63102RC & & $-$0.103 (0.128) & 0.423 &  & 0.019 (0.014)  & &  0 (0)  \\
NM016359 & & $-$0.625 (0.217) & 0.004 &  & $-$0.569 (0.143)  & &  $-$0.603 (0.170)  \\
NM002019 & & $-$0.314 (0.128) & 0.014 &  & $-$0.117 (0.076)  & &  0 (0)  \\
NM006101 & & 0.138 (0.191) & 0.471 &  & $-$0.001 (0.001)  & &  0 (0)  \\
NM014889 & & 0.276 (0.131) & 0.035 &  & 0.186 (0.095)  & &   0.429 (0.122) \\
NM004702 & & 0.967 (0.206) & $<$0.001 &  & 0.937 (0.169) & &  0.970 (0.180)  \\[6pt]
\textbf{Latency} & &  &  &  &   & &    \\
Lymph-node-positive & & $-$0.520 (0.285) & 0.068 &  & $-$0.437 (0.056)  & & $-$0.533 (0.280)  \\
Mastectomy & & $-$0.149 (0.210) & 0.479 &  & 0 (0)  & &  $-$0.134 (0.208)  \\
Chemotherapy & & 0.101 (0.285) & 0.722 &  & 0.073 (0.022)  & &  0.089 (0.284)  \\
Hormonal therapy & & $-$0.114 (0.365) & 0.755 &  & $-$0.018 (0.006)  & &  $-$0.113 (0.359)  \\
Age & & $-$0.197 (0.095) & 0.039 &  &  0.001 (0) & & $-$0.167 (0.094)  \\
Tumour diameter & & 0.323 (0.106) & 0.002 &  & 0.268 (0.057)  & &  0.332 (0.103)  \\
AA555029RC & & 0.015 (0.107) & 0.887 &  &  $-$0.002 (0) & & 0 (0)  \\
AB037863 & & 0.263 (0.114) & 0.021 &  &  0.177 (0.042) & &  0.292 (0.112)  \\
AF055033 & & $-$0.002 (0.101) & 0.982 &  &  0 (0) & &  0 (0)  \\
AK000745 & & $-$0.303 (0.120) & 0.012 &  &  $-$0.096 (0.013) & &  $-$0.270 (0.111)  \\
AF257175 & & $-$0.482 (0.120) & $<$0.001 &  &  $-$0.413 (0.045) & &  $-$0.459 (0.116)  \\
Contig46218RC & & 0.028 (0.154) & 0.855 &  & 0 (0)  & &   0 (0) \\
Contig55377RC & & $-$0.137 (0.133) & 0.304 &  &  $-$0.003 (0) & &  $-$0.084 (0.132)  \\
Contig32185RC & & $-$0.640 (0.134) & $<$0.001 &  & $-$0.515 (0.058)  & &  $-$0.621 (0.130)  \\
Contig51464RC & & 0.111 (0.118) & 0.347 &  &  0 (0) & &  0 (0)  \\
Contig63102RC & & $-$0.187 (0.107) & 0.080 &  &  $-$0.003 (0) & &  $-$0.185 (0.104)  \\
NM016359 & & 1.052 (0.199) &  $<$0.001 &  &  0.871 (0.116) & &  1.078 (0.168)  \\
NM002019 & & 0.415 (0.120) & 0.001 &  & $-$0.053 (0.010)  & &  0.427 (0.116)  \\
NM006101 & & $-$0.240 (0.165) & 0.144 &  &  $-$0.429 (0.031) & & $-$0.237 (0.149)   \\
NM014889 & & $-$0.280 (0.111) & 0.012 &  &  $-$0.241 (0.022) & &  $-$0.302 (0.107)  \\
NM004702 & & $-$0.085 (0.147) & 0.563 &  & $-$0.003 (0) & & 0 (0)   \\[6pt]
\textbf{Variance component} & &  &  &  &   & &    \\
$\theta$ & & 0.124 &  &  & 0.124  & &  0.124  \\
\hline
\end{tabular}
}
\end{threeparttable}
\end{table*}

Comparatively, the $p$-value approach based on the unpenalized model selects fewer variables than adaptive lasso and SCAD. Masud et al. (2018) noted that $p$-value can lead to erroneous exclusions of significant effects. As observed in Table~\ref{t:three}, this occurs because the unpenalized model yields inflated standard error estimates, resulting from increased collinearity due to the existence of spurious covariates.  In the incidence part of the unpenalized model (Table~\ref{t:three}), tumour diameter has a negative coefficient estimate. This suggests that an increase in tumour size increases the probability of being cured although this effect is not significant. This seems implausible in practical sense. As seen in Table~\ref{t:three} adaptive lasso shrinks this coefficient to zero and SCAD provides a relatively small positive effect. These results have good practical interpretability. For adaptive lasso and SCAD the significant variables identified do not overlap completely. Their discrepancies are largely due to one failing to exclude some small effects, while the other sets such effects to zero. For example, in the incidence submodel, adaptive lasso provides a null effect for Contig32185RC, while SCAD estimates this effect as -0.002. Applying a threshold to exclude small effects would result in greater consistency between adaptive lasso and SCAD for the results reported in Table~\ref{t:three}. 

Comparing the latency and the incidence component, we found some contrasting positive-negative coefficient estimates for certain predictor variables. These include lymph-node-positive and some gene expression variables (e.g., AB037863, NM016359, NM014889). For instance, lymph-node-positive has positive coefficient estimate in the incidence, but is negative in the latency. The incidence suggests that lymph-node-positive disease patients are less likely to be cured, compared to lymph-node-negative disease patients. However, in the latency the uncured lymph-node-positive disease patients have lower risk for breast cancer recurrence than the lymph-node-negative disease patients who are uncured.  These contrasting findings are evident in Figure~\ref{fig:fig3_KM}(B). It is apparent that lymph-node-positive disease patients had better recurrence-free survival rate than the lymph-node-negative disease patients at the early to mid-stage of follow-up. Nevertheless, the curves crossover, followed by plateaus, exhibiting a higher cure probability at the later stages of follow-up in favour of lymph-node-negative disease patients. Similar pattern may have occurred for the gene expression variables AB037863, NM016359, and NM014889.  These findings are complementary to the value of the proposed mixture-based penalized variable selection methods, and signal that a single variable may provide differing prognostic value in heterogeneous groups. 

The variance component parameter $\theta$ is estimated to be 0.124 under the unpenalized model, adaptive lasso, and SCAD because the estimator of $\theta$ is the same for these methods as discussed in Section~\ref{penREML}. We do not provide a test of statistical significance for $\theta$ because its null hypothesis lies on the boundary of parameter space. For this reason, the normal approximation of the null test statistic is inappropriate (Vaida and Xu 2000). However, we note that other approaches such as the likelihood ratio test with corrected null distribution (Vaida and Xu, 2000) or random effect prediction inference (Tawiah et al., 2019) may be used to highlight the significance of random effect. Penalized estimation methods allowing joint variable selection for fixed and random effects (e.g., Bondell et al. 2010) could also be useful in this setting. In Appendix D, we provide results (Table D1) from a penalized frailty mixture cure model with covariance structure $D(\theta)=\theta G(\rho)$. There, the results are discussed and compared with the model used in this section which is based on the covariance structure $D(\theta)=\theta I_{m}$.

\section{Conclusion}  
Due to the abundance of data sets with large volumes of candidate predictors, variable selection remains an important topic. However, to the best of our knowledge methodological studies on variable selection techniques for handling recurrent events data in the presence of a cure proportion are rare. Our development in this paper, therefore, fills this literature gap, providing methodological approaches for performing variable selection in frailty mixture cure models. The method allows a flexible choice of covariance structure for the random effect in the model, inducing either a constant or time-varying frailty term. The proposed method is based on BLUP likelihood construction in GLMM (McGilchrist 1994; McGilchrist and Yau 1995), and penalized likelihood estimation. It generalizes the penalized LSA method of Wang and Leng (2007) and the LQA sandwich variance estimator of Fan and Li (2001) to a complex setting, unifying mixture and random effect models tailored to the analysis of recurrent time-to-event data. This generalization naturally extends to a broader class of mixed models as long as a joint asymptotic covariance matrix is available for the fixed and random effects. As a semiparameteric survival variable selection method, the BLUP likelihood construction is appealing because it preserves the cancellation property of the elegant partial likelihood method. Moreover, it circumvents the intractable integration in the marginal likelihood of the model to provide an explicit form which allows ML techniques to be carried out. The explicit form is still available where the marginal likelihood is further complicated by multiple integrals due to the presence of higher order random effect terms. Hence, the proposed variable selection method can be extended to models with nested or multilevel random effects. As such, the BLUP approach is more advantageous, compared to other candidate methods such as numerical integration methods that are usually infeasible in higher order random effect settings. Also, as a semiparametric method, numerical integration may be restrictive and difficult to implement because the integrand includes a nonparameteric term.  We utilized adaptive lasso and SCAD penalty functions in the method, as noted earlier, due to their popularity and desirability. Using both penalty functions, our simulation studies shown that the proposed method have several desirable properties much in common with those of an oracle REML estimator. However, in most cases SCAD provided marginally better performance over adaptive lasso. In particular, MSEs were consistently smaller for SCAD, though at slightest margin when compared with adaptive lasso. In principle, other penalty functions could be used. Our illustration based on real data highlighted several practical benefits of the proposed penalized variable selection method, relative to a p-value based variable selection using an unpenalized model. Nevertheless, as the method is developed within the framework of mixture models potential non-identifiability issues may occur. For detailed discussion on non-identifiability of mixture models see Ng et al. (2019). Therefore, in practice one should cautiously consider the setting where mixture cure models are appropriate (e.g., see the monograph of Peng and Yu 2021). A notable consideration is utilizing large sample size, given the complexity of the model, and massive predictor space. As the method partly depends on unpenalized estimators it is not applicable to high-dimensional settings. Extension to high dimensional modeling would be a worthy goal in future research. While we assume normality for the random effect, in practice this assumption may not always hold true. Assessing the robustness of the variable selection method to misspecification of random effect distribution is beyond the scope of this paper, though is an issue of great importance, and could be considered in future studies. 

\section*{Acknowledgments}
This work was supported by a Discovery Grant from the Australian Research Council (Grant/Award Number: DP230101671).

\section*{Code availability}
R codes for implementing the methods presented in this paper are available at the following github link \url{https://github.com/rtawiah64-crt/penFMCure-reml}.

\section*{Conflict of interest}
None declared.

\section*{Appendix}

\section*{Appendix A: Time-varying AR(1) frailty mixture cure model}
From (\ref{mixture}) and (\ref{latinc}); on assuming that $u \sim N(0,\theta G(\rho))$ the resulting model is a time-varying AR(1) frailty mixture cure model. Such a model has previously been proposed and studied in Tawiah et al. (2020a). However, we note that the model considered here is slightly different because the random effect $u$ is shared by the incidence, and the latency submodels, unlike the existing work (Tawiah et al. 2020a), wherein separate random effects are used. Therefore, the estimation algorithms given in Tawiah et al. (2020a) are not directly applicable to the model considered here. We take $G(\rho)$ to be an $n \times n$ block diagonal matrix given by  
\[
G(\rho)=\begin{pmatrix}
G_{1}(\rho) & 0 & \dots & 0 \\
& G_{2}(\rho) & \dots & 0 \\
& \multirow{2}{*}{\makebox[0pt]{\text{%
Sym.}}} & \ddots & \vdots  \\
&
&
& G_{m}(\rho)
\end{pmatrix}
\hspace{0.3cm}\mbox{and} \hspace{0.3cm}
G_{i}(\rho)=\frac{1}{1-\rho^{2}}\begin{pmatrix}
1 & & &\rho & \dots & \rho^{n_{i}-1}  \\
& & & 1 & \dots & \rho^{n_{i}-2} \\
& & & \multirow{2}{*}{\makebox[0pt]{\text{%
Sym.}}}  & \ddots & \vdots  \\
&
&
& & & 1
\end{pmatrix},
\] 
where Sym. indicates that the matrix is symmetric, $G_{i}(\rho)$ is an AR(1) covariance matrix for subject $i$. In the time-varying frailty mixture cure model, $\alpha$ and $\beta$ are estimated as discussed in Section \ref{meth} for the constant frailty mixture cure model. However, the variance component parameters $\rho$ and $\theta$ are estimated differently using REML. The REML estimating equation for $\theta$ is given by    
\begin{equation}\tag{A.1}
\theta =n^{-1}\left( \mbox{tr} G^{-1}\Sigma_{u,u}^{*} +u^{T}G^{-1}u\right), 
\end{equation}
where $G^{-1}=\left(1+\rho^{2} \right) I-\rho J-\rho^{2}K$. Note that $I, J$ and $K$ are block diagonal matrices with respective sub-entries $I_{i}, J_{i}$ and $K_{i}$ each with $n_{i}$ rows and columns, given by
\[
I_{i}=
\begin{bmatrix}
1 & &0 & \cdots & 0\\
0 & & 1 & \cdots & 0\\
\vdots & & \vdots & \ddots & \vdots\\
0 & & 0 & \cdots & 1\\
\end{bmatrix},  \hspace{0.3cm}
J_{i}=
\begin{bmatrix}
0 & & 1 & \cdots & 1\\
0 & & 0 & \cdots & 1\\
\vdots & & \vdots & \ddots & \vdots\\
1 & & 1 & \cdots & 0\\
\end{bmatrix} \hspace{0.3cm}
\mbox{and}  \hspace{0.3cm}
K_{i}=
\begin{bmatrix}
1 & & 0 & \cdots & 0\\
0 & & 0 & \cdots & 0\\
\vdots & & \vdots & \ddots & \vdots\\
0 & & 0 & \cdots & 1\\
\end{bmatrix} .
\] 
Substituting the expresion of $G^{-1}$ into (A.1), the estimating equation of $\theta$ reduces to
\begin{equation}\tag{A.2} 
\theta=n^{-1}\left\lbrace  \left(1+\rho^{2} \right)B_{1} -2\rho B_{2}-\rho^{2}B_{3} \right\rbrace , 
\end{equation}
where $  B_{1}=\mbox{tr}\left\lbrace  I \left(\Sigma_{u,u}^{*}+uu^{T} \right)\right\rbrace  , B_{2}=\left( 1/2\right)\mbox{tr}\left\lbrace J\left( \Sigma_{u,u}^{*}+uu^{T}\right) \right\rbrace $ and\\ $B_{3}=\mbox{tr}\left\lbrace K \left( \Sigma_{u,u}^{*}+uu^{T}\right) \right\rbrace $. 
The REML estimating equation for $\rho$  is a solution to the  equation
\begin{equation}\tag{A.3}
\mbox{tr} \left\lbrace  \left(  \frac{\partial G^{-1}}{\partial \rho}\right) G \right\rbrace   =\frac{1}{\theta}\left\lbrace  \mbox{tr} \left(G+uu^{T} \right)\frac{\partial G^{-1}}{\partial \rho} \right\rbrace ,  
\end{equation}
where $ \mbox{tr} \left((\partial G^{-1}/ \partial \rho) G \right)=- \left[2 \rho /\left(1-\rho^{2} \right)  \right] $, and $ \partial G^{-1}/ \partial \rho=2\rho I - J-2 \rho K$ (Yau and McGilchrist 1998). Substituting these expressions into (A.3) and replacing $\theta$ by its REML equation (A.2), we obtain the cubic function
\[
f\left(\rho \right) =A_{1}\rho^{3}+A_{2}\rho^{2}+A_{3}\rho+A_{4}=0,
\]
where $A_{1}=(n-m)(B_{1}-B_{3}), A_{2}=(2m-n)B_{2}, A_{3}=nB_{3}-(n+m)B_{1}$, and $A_{4}=nB_{2}$. The estimate of $\rho$ is found by a numerical search through a Newton-Raphson solution of the above cubic equation given by
\begin{equation}\tag{A.4}
\hat{\rho}=\rho-\frac{f(\rho)}{f'(\rho)},
\end{equation}
where $f'(\rho)$ is the first derivative of $f(\rho)$ with respect to $\rho$. Note that once $\alpha$, $\beta$, $\theta$, and $\rho$ are calculated the corresponding penalized fixed effect estimates $\hat{\alpha}_{\kappa_1}=(\hat{\alpha}_{0},\hat{\alpha}_{\kappa_1,1},...,\hat{\alpha}_{\kappa_1,d})$ and
$\hat{\beta}_{\kappa_2}= (\hat{\beta}_{\kappa_2,1},...,\hat{\beta}_{\kappa_2,p})$ are produced by maximizing the approximate “augmented” data penalized BLUP log-likelihood (\ref{apenblup}), and performing regularization in the manner presented in Section~\ref{tuning}. 

\section*{Appendix B: Supplementary details of the BLUP method}
Let $g$, $\xi$, and $\eta$  denote the vectors of $g_{ij}$, $\xi_{ij}$, and $\eta_{ij}$, respectively. The derivation of $\partial \ell_{1}/\partial \xi$ and $\partial^2 \ell_{1}/\partial \xi\partial \xi^T$ are given by
\[
\frac{\partial \ell_{1}}{\partial \xi}=g-\frac{\exp(\xi)}{1+\exp(\xi)}
\hspace{0.2cm} \mbox{and} \hspace{0.2cm}
-\frac{\partial ^{2}\ell_{1}}{\partial \xi \partial \xi ^{T}}=\mbox{diag}\left\lbrace \frac{\exp(\xi)}{\left( 1+\exp(\xi)\right) ^{2}}
\right\rbrace , 
\] 
pertaining to the incidence submodel. For the latency submodel, we let $\varpi_{b}=g_{b}exp\left(\eta_{b} \right) $, $e_{b}=\delta_{b}/\sum_{j=1}^{N}\varpi_{j}$, $s_{q}=\sum_{j=1}^{N}e_{j}$, $Q=diag(\varpi_{1},...,\varpi_{N})$, $E=diag(e_{1},...,e_{N})$, $S=(s_{1},...,S_{N})$, $1=(1,...,1)^T$ and 
$F_{1}=\begin{pmatrix}
1 & \dots & 0\\
\vdots & \ddots & \vdots\\
1 & \dots & 1\\
\end{pmatrix}$. It follows that
\[
\frac{\partial \ell_{1}}{\partial \eta}=\Delta-QFE1
\hspace{0.2cm} \mbox{and} \hspace{0.2cm}
-\frac{\partial^{2}\ell_{1}}{\partial \eta \partial \eta^{T}}=QS-QFE^{2}F^{T}Q,
\]
where $\Delta$ is the censoring vector. The matrix of the negative second derivatives  $\Sigma=-\partial^2 \ell(\Omega,\theta)/\partial\Omega\partial\Omega^2$ is given by
\[\Sigma=
\begin{bmatrix}
W^T & & 0\\
0 & & Z^T\\
R^T & &  R^T\\
\end{bmatrix}
\begin{bmatrix}
\Gamma_{\xi\xi} & & \Gamma_{\xi\eta}\\
\Gamma_{\eta\xi} & & \Gamma_{\eta\eta}\\
\end{bmatrix}
\begin{bmatrix}
W & & 0 & & R \\
0 & & Z & & R \\
\end{bmatrix}+
\begin{bmatrix}
0 & & 0 & & 0 \\
0 & & 0 & & 0 \\
0 & & 0 & & D(\theta)^{-1}\\
\end{bmatrix},
\]
where 
\[
\Gamma_{\xi\xi} = -\frac{\partial ^{2}\ell_{1}}{\partial \xi \partial \xi ^{T}}, \hspace{0.3cm}
\Gamma_{\eta\eta} = -\frac{\partial^{2}\ell_{1}}{\partial \eta \partial \eta^{T}}, \hspace{0.3cm} \mbox{and}
\hspace{0.3cm}
\Gamma_{\xi\eta} = \Gamma_{\eta\xi}=0.
\]
Note that $D(\theta)^{-1}=\theta^{-1}I_{m}$ where $u\sim N(0, \theta I_{m})$, provides a constant frailty mixture cure model. On the other hand, $D(\theta)^{-1}=\theta^{-1}G^{-1}$ where $u\sim N(0,\theta G(\rho))$, yields a time-varying AR(1) frailty mixture cure model. The matrix $I_{m}$ is defined in Section~\ref{meth}, and $G(\rho)$ is given in Appendix A.

\section*{Appendix C: Covariance of penalized estimators}
Considering the LQA sandwich formula (Fan and Li 2001), we define the covariance matrix of $\hat{\Omega}_{\kappa}$ as
\begin{equation}\tag{C.1}
\widehat{\mbox{cov}}(\hat{\Omega}_{\kappa})=\left\lbrace Q''(\hat{\Omega}_{\kappa},\hat{\theta})+n\Psi_{\kappa}(\hat{\Omega}_{\kappa})\right\rbrace^{*}\widehat{\mbox{cov}}\left\lbrace  Q'(\hat{\Omega}_{\kappa},\hat{\theta})\right\rbrace \left\lbrace Q''(\hat{\Omega}_{\kappa},\hat{\theta})+n\Psi_{\kappa}(\hat{\Omega}_{\kappa})\right\rbrace^{*},
\end{equation}
where $Q'(\hat{\Omega}_{\kappa},\hat{\theta})$ and $Q''(\hat{\Omega}_{\kappa},\hat{\theta})$ are the first and second derivatives of $Q(\Omega,\theta) $ at $\hat{\Omega}_{\kappa}$. The matrix $\Psi_{\kappa}(\hat{\Omega}_{\kappa})$ is defined by 
\[
\Psi_{\kappa}(\hat{\Omega}_{\kappa})=\mbox{diag}\left\lbrace \frac{\varphi'_{\kappa_1}(|\alpha_{10}|)}{|\alpha_{10}|},...,
\frac{\varphi'_{\kappa_1}(|\alpha_{d0}|)}{|\alpha_{d0}|},     \frac{\varphi'_{\kappa_2}(|\beta_{10}|)}{|\beta_{10}|},...,
\frac{\varphi'_{\kappa_2}(|\beta_{10}|)}{|\beta_{p0}|}, 0_{10},...,0_{m0} \right\rbrace,
\]
and
\[
\widehat{\mbox{cov}}\left\lbrace  Q'(\hat{\Omega}_{\kappa},\hat{\theta})\right\rbrace  = \left\lbrace Q''(\hat{\Omega}_{\kappa},\hat{\theta})+A(\hat{\Omega}_{\kappa}) \right\rbrace \left\lbrace Q''(\hat{\Omega}_{\kappa},\hat{\theta})\right\rbrace^{*}  \left\lbrace Q''(\hat{\Omega}_{\kappa},\hat{\theta})+A(\hat{\Omega}_{\kappa}) \right\rbrace ,
\]
where $A(\hat{\Omega}_{\kappa})$ is   
\[
A(\hat{\Omega}_{\kappa})= \mbox{diag}\left\lbrace \frac{I(\alpha_{0}\neq 0)}{\alpha_{0}^2},...,
\frac{I(\alpha_{d}\neq 0)}{\alpha_{d}^2},
\frac{I(\beta_{1}\neq 0)}{\beta_{1}^2},...,
\frac{I(\beta_{p}\neq 0)}{\beta_{p}^2}, 0_{1},...,0_{m} \right\rbrace .
\]
As discussed in Section~\ref{meth}, we let $\left\lbrace Q''(\hat{\Omega}_{\kappa},\hat{\theta})+n\Psi_{\kappa}(\hat{\Omega}_{\kappa})\right\rbrace^{*}$ to denote the Cholesky decomposition of  $\left\lbrace Q''(\hat{\Omega}_{\kappa},\hat{\theta})+n\Psi_{\kappa}(\hat{\Omega}_{\kappa})\right\rbrace$, where $Q''(\hat{\Omega}_{\kappa},\hat{\theta})$ is replaced by $\Sigma$.

\section*{Appendix D: Data application using the time-varying AR(1) frailty mixture cure model}
To allow comparison, we applied the time-varying AR(1) frailty mixture cure model $\left[ i.e., u\sim N(0,\theta G(\rho))\right] $ to the breast cancer gene expression data, using an unpenalized REML, adaptive lasso, and SCAD penalized methods for estimation. Results are reported in Table D1. The tuning parameters selected by adaptive lasso and SCAD are $\kappa =(0.009, 0.051)$ and $\kappa = (0.018, 0.045)$, respectively.
\begin{table*}[!htb]
\centering
\renewcommand\thetable{D1}
\caption*{Table D1: Unpenalised REML method and the proposed penalised REML method with adaptive lasso and SCAD penalty applied to the breast cancer data. Underlying model is the AR(1) frailty mixture cure model.}
\begin{threeparttable}
\resizebox{13.5cm}{!}{
\begin{tabular}{lccccccc}
\hline
&  & \multicolumn{2}{c}{REML}  & \multicolumn{2}{c}{Adaptive lasso} & \multicolumn{2}{c}{SCAD}    \\ 
  \cline{3-4} \cline{6-6} \cline{8-8}
 Variable  & & Estimate (ASE)  & $p$-value &  & Estimate (ASE) &  & Estimate (ASE)  \\ \hline
\textbf{Incidence} & &  &  &  &   & &    \\
Intercept & & 0.039 (0.199) & 0.845 &  &  0.081 (0.020) & &  0.073 (0.165)  \\
Lymph-node-positive & & 0.978 (0.356)& 0.006 &  & 0.971 (0.184)  & & 0.978 (0.350)  \\
Mastectomy & & $-$0.059 (0.250) & 0.812 &  &  0.001 (0) & &   0 (0) \\
Chemotherapy & & $-$0.585 (0.358) & 0.102  &  &  $-$0.579 (0.067) & &  $-$0.587 (0.352)  \\
Hormonal therapy & & $-$0.207 (0.395) & 0.600 &  & $-$0.185 (0.044)  & &  $-$203 (0.387)  \\
Age & & $-$0.212 (0.124) & 0.087 &  & $-$0.252 (0.044)  & & $-$0.215 (0.122)  \\
Tumour diameter & & $-$0.123 (0.127) & 0.332 &  & 0 (0)  & & $-$0.184 (0.121)   \\
AA555029RC & & $-$0.097 (0.140) & 0.489 &  & 0 (0)  & & 0 (0)  \\
AB037863 & &$-$0.244 (0.135)  & 0.069 &  &  $-$0.201 (0.042) & &  $-$0.239 (0.133)  \\
AF055033 & & 0.426 (0.137) & 0.002 &  & 0.443 (0.111)  & &  0.435 (0.133)  \\
AK000745 & & 0.243 (0.155) & 0.116 &  & 0.080 (0.038)  & &  0.248 (0.137)  \\
AF257175 & & $-$0.130 (0.138) & 0.346 &  & $-$0.117 (0.016)  & &  $-$0.188 (0.127)  \\
Contig46218RC & & 0.033 (0.189) & 0.860 &  & 0 (0)  & &  0 (0)  \\
Contig55377RC & & 0.253 (0.155) & 0.103 &  & 0.242 (0.078)  & &  0.246 (0.146)  \\
Contig32185RC & & $-$0.024 (0.157) & 0.879 &  & 0 (0)  & &  0 (0)  \\
Contig51464RC & & $-$0.077 (0.137) & 0.576 &  & $-$0.001 (0)  & &  0 (0)  \\
Contig63102RC & & $-$0.156 (0.128) & 0.224 &  &  $-$0.110 (0.018) & &  $-$0.153 (0.126)  \\
NM016359 & & $-$0.261 (0.214) & 0.221 &  &  $-$0.321 (0.051) & &  $-$0.260 (0.174)  \\
NM002019 & & $-$0.189 (0.128) & 0.140 &  & 0.047 (0.012)  & & $-$0.175 (0.121)   \\
NM006101 & & 0.061 (0.189) & 0.745 &  & $-$0.007 (0.001)  & & 0 (0)   \\
NM014889 & & 0.131 (0.128) & 0.307 &  & 0.002 (0.001)  & & 0.113 (0.123)   \\
NM004702 & & 0.761 (0.202) & $<$0.001 &  & 0.555 (0.125) & &  0.783 (0.186)  \\[6pt]
\textbf{Latency} & &  &  &  &   & &    \\
Lymph-node-positive & & $-$0.490 (0.303) & 0.105 &  &  $-$0.414 (0.082) & & $-$0.518 (0.214)  \\
Mastectomy & & $-$0.116 (0.228) & 0.612 &  & 0.002 (0.001)  & & $-$0.012 (0.001)   \\
Chemotherapy & & 0.047 (0.303) & 0.877 &  & 0 (0)  & &   0 (0) \\
Hormonal therapy & & $-$0.031 (0.397) & 0.937 &  & 0 (0)  & &  0 (0)  \\
Age & & $-$0.211 (0.104) & 0.042 &  & $-$0.374 (0.011)  & &  0 (0) \\
Tumour diameter & & 0.385 (0.118) & 0.001 &  & 0.139 (0.031)  & &  0.434 (0.110)  \\
AA555029RC & & 0.183 (0.116) & 0.114 &  & 0.042 (0.008)  & & 0.164 (0.107)  \\
AB037863 & & 0.126 (0.127) & 0.319 &  & $-$0.008 (0.001)  & &  0.007 (0.004)  \\
AF055033 & & 0.129 (0.106) & 0.221 &  & 0.001 (0)  & &  0 (0)  \\
AK000745 & & $-$0.400 (0.130) & 0.002 &  &  $-$0.374 (0.023) & &  $-$0.388 (0.200)  \\
AF257175 & & $-$0.300 (0.131) & 0.022 &  & $-$0.214 (0.012)  & &  $-$0.256 (0.125)  \\
Contig46218RC & & 0.109 (0.163) & 0.503 &  & $-$0.053 (0.007)  & & 0.004 (0.002)   \\
Contig55377RC & & $-$0.118 (0.147) & 0.420 &  & 0 (0)  & &  $-$0.010 (0.003)  \\
Contig32185RC & & $-$0.577 (0.141) & $<$0.001 &  & $-$0.416 (0.031)  & &  $-$0.554 (0.131)  \\
Contig51464RC & & 0.208 (0.125) & 0.096 &  & 0.205 (0.036)  & &  0.318 (0.118)  \\
Contig63102RC & & $-$0.188 (0.118) & 0.113 &  & $-$0.024 (0.002)  & &   0 (0) \\
NM016359 & & 0.805 (0.215) & $<$0.001 &  &  0.854 (0.130) & &   0.831 (0.178) \\
NM002019 & & 0.270 (0.126) & 0.032 &  & 0.327 (0.057)  & &  0.318 (0.124)  \\
NM006101 & & $-$0.223 (0.181) & 0.218 &  & $-$0.134 (0.016)  & & $-$0.236 (0.152)   \\
NM014889 & & $-$0.198 (0.120) & 0.099 &  & $-$0.038 (0.002)  & & $-$0.365 (0.113)   \\
NM004702 & & 0.100 (0.155) & 0.518 &  & 0 (0) & & 0.002 (0.001)   \\[6pt]
\textbf{Variance component} & &  &  &  &   & &    \\
$\theta$ & & 0.077 &  &  &  0.077 & &  0.077  \\
$\rho$ & & $-$0.952 &  &  & $-$0.952  & &  $-$0.952  \\
\hline
\end{tabular}
}
\end{threeparttable}
\end{table*}
 
Comparing with the constant frailty cure model (Table 3 in Section~\ref{appl}), we observe some agreements, but also differences in the results provided by the variable selection methods. For instance, in the constant frailty cure model SCAD identifies age as an insignificant predictor for cure probability, but it selects age as a significant variable in the AR(1) frailty cure model. Likewise, based on $p$-values NM0163593 and NM002019 are significant in the incidence of the unpenalized constant frailty cure model, but they are insignificant in the unpenalized AR(1) frailty cure model. In fact, this  is  not surprising because, different covariance structures are expected to produce slightly different results. Therefore, users should be well informed about the covariance structure that best suits their data sets. This may be based on domain knowledge about the characteristics of the data. The AR(1) covariance structure allows time-varying frailty. It is most useful if the occurrence of initial events modify (i.e., increase or decrease) the hazard function for the next event.   For example, in heart disease studies a one or more previous episodes of myocardial infarction may deteriorate the heart muscle over time and increase the hazard for recurrence or reduce the cure probability. However, in some situations the hazard function does not change over time and so the constant frailty appears to be a reasonable choice. An example is the set of data from the chronic granulomatous disease (CGD) trial (Fleming and Harrington, 1991) which has been used widely to illustrate the methodology of recurrent events data. As pointed out in Therneau and Grambsch (2000) the clinical scientists who conducted the trial learned from practical experience that the risk of recurrent infections in CGD patients remained constant regardless of the number of previous infections. Alternatively, there are formal statistical approaches to selecting an appropriate covariance structure in GLMMs (e.g., see Cai and Dunson, 2006).

\end{document}